\documentclass[structabstract]{aa}  
\usepackage{multirow} 
\usepackage{natbib}
\usepackage{amssymb,amsmath}
\usepackage{graphicx, subfigure}
\usepackage{rotating}
\usepackage{txfonts}
\usepackage[cmyk]{xcolor} 
\newcommand{\mdot}{\ensuremath{\dot{M}}}  

\begin{document}

   \title{The wind of W\,Hya as seen by Herschel\thanks{Herschel is an ESA space observatory with science instruments provided by
European-led Principal Investigator consortia and with important participation
from NASA.}}

   \subtitle{I. The CO envelope}

   \author{T. Khouri
          \inst{1}\thanks{{\it Send offprint requests to T. Khouri}\newline \email{theokhouri@gmail.com}}, A. de Koter \inst{1,2}, L. Decin\inst{1,2}, L. B. F. M. Waters \inst{1,3},
           R. Lombaert \inst{2}, P. Royer \inst{2}, B. Swinyard \inst{4,5}, M. J. Barlow \inst{5}, J. Alcolea\inst{6}, J. A. D. L. Blommaert\inst{2,7}, V. Bujarrabal \inst{8}, J. Cernicharo \inst{9},
           M. A. T. Groenewegen \inst{10}, K. Justtanont \inst{11}, F. Kerschbaum \inst{12}, M. Maercker \inst{13}, A. Marston\inst{14}, M. Matsuura\inst{5}, G. Melnick\inst{15}, K. M. Menten \inst{16}, H. Olofsson \inst{11,17},
           P. Planesas \inst{6}, E. Polehampton \inst{4,18}, Th. Posch \inst{12}, M. Schmidt \inst{19}, R. Szczerba \inst{19}, B. Vandenbussche\inst{2}, J. Yates \inst{5}
          }

   \institute{Astronomical Institute ÒAnton PannekoekÓ, University of Amsterdam, PO Box 94249, 1090 GE Amsterdam, The Netherlands
         \and
            Instituut voor Sterrenkunde, KU Leuven, Celestijnenlaan 200D B-2401, 3001 Leuven, Belgium
            \and
            SRON Netherlands Institute for Space Research, Sorbonnelaan 2, 3584 CA Utrecht, The Netherlands
            \and
            RAL Space, Rutherford Appleton Laboratory, Chilton, Didcot, Oxfordshire, OX11 0QX, UK
            \and
            Dept. of Physics \& Astronomy, University College London, Gower St, London WC1E 6BT, UK
            \and
            Observatorio Astron\'omico Nacional (IGN), Alfonso XII N$^\circ$3, 28014 Madrid, Spain
            \and
            Department of Physics and Astrophysics, Vrije Universiteit Brussel, Pleinlaan 2, 1050 Brussels, Belgium
            \and
            Observatorio Astron\'omico Nacional (OAN-IGN), Apartado 112, E-28803 Alcal\'a de Henares, Spain
            \and
            Centro de Astrobiolog\'ia (CSIC/INTA), Ctra. de Torrej\'on a Ajalvir, km 4, 28850 Torrej\'on de Ardoz, Madrid, Spain
            \and
            Koninklijke Sterrenwacht van Belgi\"e, Ringlaan 3, B--1180 Brussel, Belgium
            \and
           Onsala Space Observatory, Dept. of Earth and Space Sciences, Chalmers University of Technology, SE-43992 Onsala, Sweden
           \and
            University of Vienna, Department of Astrophysics, T\"urkenschanzstra\ss e 17, 1180 Wien, Austria
            \and
            European Southern Observatory, Karl Schwarzschild Str. 2, 85748 Garching bei M\"unchen, Germany
            \and
            European Space Astronomy Centre, Urb. Villafranca del Castillo, PO Box 50727, 28080 Madrid, Spain
            \and
           Harvard-Smithsonian Center for Astrophysics, Cambridge, MA 02138, USA
            \and
            Max-Planck-Institut f\"ur Radioastronomie, Auf dem H\"ugel 69, 53121 Bonn, Germany
            \and
            Department of Astronomy, AlbaNova University Center, Stockholm University, 10691 Stockholm, Sweden
            \and
            Institute for Space Imaging Science, University of Lethbridge, 4401 University Drive, Lethbridge, Alberta, T1J 1B1, Canada
           \and
            N. Copernicus Astronomical Center, Rabia\'nska 8, 87-100 Toru\'n, Poland
             }


  \abstract
   {Asymptotic giant branch (AGB) stars lose their envelopes by means of a stellar wind whose driving mechanism is not understood well.
   Characterizing the composition and thermal and dynamical structure of the outflow provides constraints that are essential for understanding AGB
   evolution, including the rate of mass loss and isotopic ratios.}
   {We characterize the CO emission from the wind of the low mass-loss rate oxygen-rich AGB star W\,Hya using data obtained by the HIFI, PACS, and SPIRE instruments onboard the Herschel
   Space Observatory and ground-based telescopes. $^{12}$CO and $^{13}$CO lines are used to constrain the intrinsic $^{12}$C/$^{13}$C  ratio from resolved HIFI lines.}
   {
   We combined a state-of-the-art molecular line emission code and a dust continuum radiative transfer code to model the CO lines and the thermal dust continuum.
   }
   {The acceleration of the outflow up to about 5.5 km/s is quite slow and can be represented by a $\beta$-type velocity law with index $\beta = 5$. Beyond this point,
   acceleration up the terminal velocity of 7 km/s is faster. Using the $J$= 10--9, 9--8, and 6--5 transitions, we find an intrinsic $^{12}$C/$^{13}$C ratio of $18 \pm 10$
   for W\,Hya, where the error bar is mostly due to uncertainties in the $^{12}$CO
abundance and the stellar flux around 4.6 $\mu$m. 
To match the low-excitation CO lines, these molecules need to be photo-dissociated at $\sim$\,500 stellar radii.
The radial dust emission intensity profile of our stellar wind model matches PACS images at 70 $\mu$m out to 20$\arcsec$ (or 800 stellar radii). 
For larger radii the observed emission is substantially stronger than our model predicts, indicating that at these locations there is extra material present. 
}
   {The initial slow acceleration of the wind may imply inefficient dust formation or dust driving in the lower part of the envelope.
   The final injection of momentum in the wind might be the result of an increase in the opacity thanks to the late condensation of dust species.
The derived intrinsic isotopologue ratio for W\,Hya is consistent with values set by the first dredge-up and suggestive of an initial mass of 2 M$_\odot$ or more.
However, the uncertainty in the isotopologic ratio is large, which makes it difficult to set reliable limits on W\,Hya's main-sequence mass.
}

               
\titlerunning{The CO envelope of W\,Hya}
\authorrunning{Khouri et al.}

\maketitle
%

\section{Introduction}

The asymptotic giant branch (AGB) represents one of the final evolutionary stages of low and intermediate mass stars.
AGB objects are luminous and have very extended, weakly gravitationally bound and cool atmospheres. Their outermost
layers are expelled by means of a dusty stellar wind \citep[e.g.][]{Habing2003}. The high mass-loss rate during the AGB
phase prevents stars with masses between 2 M$_{\odot}$ and 9 M$_{\odot}$ from evolving to the supernova stage.

During their lives, low and intermediate mass stars may undergo three distinct surface enrichment episodes. Those are referred to
as dredge-ups, and they happen when the convective streams from the outer layers reach deep into the interior.
Elements synthesized by nuclear fusion or by slow-neutron capture in the interior of AGB stars are brought to the surface and eventually ejected
in the wind \citep[e.g.][]{Habing2003}. In this way, AGB stars contribute to the chemical enrichment of the interstellar medium and, in
a bigger context, to the chemical evolution of galaxies.

First and second dredge-up processes occur when these stars ascend the giant branch \citep{Iben1983}.
The third dredge-up is in fact a series of mixing events during the AGB phase, induced by thermal pulses
\citep[TPs,][]{Iben1975}. Evolutionary models predict by how much the surface abundance of each element is enriched for a star with a given initial
mass and metallicity \citep[][and references therein]{Ventura2009, Karakas2010, Cristallo2011}. These models, however,
need to be compared with observations. In particular, the change in surface chemical composition
of AGB stars depends on initial mass and the assumed mass loss as a function of time. A very powerful diagnostic for the enrichment processes are surface
isotopic ratios. From evolutionary model calculations, these ratios are found to vary strongly depending on the
dredge-up events the star has experienced, and, therefore, evolutionary phase, and on the main sequence mass of the star \citep[e.g.][and references therein]{Boothroyd1999,Busso1999,Charbonnel2010,Karakas2011}.

Since the outflowing gas is molecular up to large distances from the star (typically up to $\approx$ 1000 R$_{\star}$),
the isotopic ratios must be retrieved from isotopologic abundance ratios.
In this paper, we use an unparalleled number of $^{12}$CO and $^{13}$CO emission lines and the thermal infrared
continuum to constrain the structure of the outflowing envelope of the oxygen-rich AGB star W Hya and, specifically,
the isotopic ratio $^{12}$C/$^{13}$C.

W\,Hya was observed by the three
instruments onboard the Herschel Space Observatory
\citep[hereafter {\it Herschel};][]{Pilbratt2010}. These are the Heterodyne Instrument for the Far Infrared, HIFI, \citep{deGraauw2010}, the Spectral and Photometric Imaging Receiver Fourier-Transform Spectrometer,
SPIRE FTS, \citep{Griffin2010}, and the Photodetector Array Camera and Spectrometer, PACS, \citep{Poglitsch2010}. We supplemented these data with earlier observations from ground-based
telescopes and the Infrared Space Observatory \citep[ISO;][]{Kessler1996}. The observations
carried out by {\it Herschel} span an unprecedented range in excitation energies for the ground vibrational level and cover, in the case of W\,Hya, CO lines from an upper rotational level $J_{\rm up}$= 4 to 30.
When complemented with
ground-based observations of lower excitation transitions, this dataset offers an unique picture of the outflowing molecular envelope of W\,Hya. This allows us to
reconstruct the flow from the onset of wind acceleration out to the region where CO is dissociated, which is essential for understanding the poorly understood
wind-driving mechanism. 
Specifically, the velocity information contained in the line shapes of the HIFI high-excitation lines of $^{12}$CO, $J$=16-15 and 10-9 probe the acceleration in the inner part of the flow; the
integrated line fluxes from $J$=4-3 to $J$=11-10 measured by SPIRE and the highest $J$ transitions observed by PACS give a very complete picture of the CO excitation throughout the wind.
Finally, the low-$J$ transitions secured from the ground probe the outer regions of the flow. 

The availability of multiple $^{13}$CO transitions, in principle, permits constraints to be placed on the intrinsic  $^{13}$CO/$^{12}$CO isotopologic ratio. A robust determination of this
ratio is sensitive to stellar parameters and envelope properties, and for this reason is not easy to obtain. We modelled the CO envelope of W\,Hya in detail
and discuss the effect of uncertainties on the parameters adopted to the derived $^{12}$CO/$^{13}$CO ratio.

Together with the gas, we simultaneously and consistently model the solid state component in the outflow and constrain the abundance and chemical properties of the
dust grains by fitting the ISO spectrum of W\,Hya.  In this way we can constrain the dust-to-gas ratio.

In Sect. \ref{sec:WHya}, the target, W\,Hya, and the available dataset are introduced, and the envelope model assumptions are presented.
Section \ref{sec:radiation_field} is devoted to discussing radiative
transfer effects that hamper determinations of isotopic ratios from line strength ratios, especially for low mass-loss rate objects.
We present and discuss the results of our models in Sects. \ref{sec:CO_WHya} and \ref{sec:disc}.
Finally, we present a summary of the points addressed in this work in Sect. \ref{sec:summary}.

\section{Dataset and model assumptions}
\label{sec:WHya}

\subsection{Basic information on W\,Hya}

W\,Hya is one of the brightest infrared sources in the sky and the second brightest AGB star in the K band \citep{Wing1971}.
This bright oxygen-rich AGB star is relatively close, but some uncertainty on its distance still exists. 
Distances reported in the literature range from 78 pc to 115 pc \citep{Knapp2003, Glass2007, 1997ESASP1200.....P}.
In this study we adopt the distance estimated by \citeauthor{Knapp2003} (78 pc), following \cite{Justtanont2005}. The current mass-loss rate of the star is fairly modest.
Estimates of this property do, however, show a broad range, from a few times 10$^{-7}$ M$_\odot$yr$^{-1}$ \citep{Justtanont2005} to a few times 10$^{-6}$ M$_\odot$yr$^{-1}$ \citep{Zubko2000}.
This range is likely the result of the use of different diagnostics (SED fitting, H$_2$O and CO line modeling), and/or model assumptions.
W\,Hya is usually classified as a semi-regular variable, although this classification is somewhat controversial \citep{Uttenthaler2011}.
Its visual magnitude changes between six and ten with a period of about 380 days.

The star was one of the first AGB stars for which observations of H$_2$O rotational
emission were reported, using the
ISO \citep{Neufeld1996,Barlow1996}.
W\,Hya was also observed
in the 557 GHz ground-state water transition using SWAS \citep{Harwit2002} and using Odin \citep{Justtanont2005}.
The water emission from this object is strong, and a lot of effort
has been invested in modelling and interpreting the observed lines \citep[e.g.][]{Neufeld1996, Barlow1996, Zubko2000,
Justtanont2005, Maercker2008, Maercker2009}. The results usually point to high water abundance
relative to H$_2$, ranging from $10^{-4}$ to a few times $10^{-3}$.

The dust envelope of W\,Hya was imaged using the Infrared Astronomical Satellite (IRAS) and found to be
unexpectedly extended \citep{Hawkins1990}.
\cite{Cox2012} reported W\,Hya to be the only oxygen-rich AGB star in their sample to show a ring-like structure.
The data, however, have to be studied 
in more detail before a firm conclusion on the cause of the structures can be drawn.

\cite{Zhao-Geisler2011} monitored W\,Hya in the near-IR (8-12 $\mu$m) with MIDI/VLTI and fitted the visibility data with a
fully limb-darkened disk with a radius of 40 mas (4 AU). The authors set a lower limit on the silicate dust shell radius of 28 photospheric radii
($50$ AU). They propose that there is a much smaller Al$_2$O$_3$ shell that causes, together with H$_2$O
molecules, the observed increase in diameter at wavelengths longer than $10\ \mu$m. This is consistent with the later observations
by \cite{Norris2012} of close-in transparent large iron-free grains. These two works point to a picture in which an inner shell with
large transparent grains is enclosed by an outer shell of more opaque silicate grains.

Furthermore, \cite{Zhao-Geisler2011} discuss evidence seen on different scales and presented by
different authors that point to a non-spherical symmetrical envelope of W\,Hya.
The observations reported in the literature usually indicate that the departures from spherical
symmetry can be explained by an ellipsoid source. The PACS 70 $\mu$m images
published recently by \cite{Cox2012} also show signs of asymmetry.

\subsection{Dataset}

We compiled a dataset of W\,Hya's $^{12}$CO and $^{13}$CO emission lines that comprises observations carried out by all instruments
onboard {\it Herschel}, as well as data from ISO and ground-based telescopes. These ground-based observatories are the Atacama Pathfinder EXperiment, APEX,
the Arizona Radio Observatory Sub-Millimeter Telescope, SMT, and the Swedish-ESO 15 m Submillimeter Telescope, SEST.

The CO lines were obtained as part of several projects:  the HIFI observations were part of the HIFISTARS {\it Herschel} guaranteed time key programme
and were presented by \cite{Justtanont2012}; the SPIRE and PACS data were obtained by the MESS consortium \citep{Groenewegen2011};
and the ground-based data were presented by \cite{DeBeck2010}.

The line properties and the measured integrated flux are listed in Table \ref{table:WHya_line_fluxes}.
The $^{13}$CO lines from W\,Hya are too weak to be seen above the noise in both the PACS and SPIRE spectra.
We did not include the integrated line fluxes for the ground-based observations (for which the source size is comparable to
the beam size of the telescopes, also presented in Table \ref{table:WHya_line_fluxes}), since the conversion from antennae temperatures to fluxes for
partially or fully spatially resolved lines is a function of the size of the emitting region.
All together, the observed transitions used to constrain our models
span a range from $J$=1-0 to $J$=24-23. This covers excitation energies of the upper level
from 5.5 K to 1656 K. We did not include lines with upper rotational levels higher than 24 in our analysis.
Transitions $J$=25-24 and $J$=26-25 are in a region where spectral leakage occurs so the measurements of PACS are not reliable.
The even higher excitation lines ($T_{\rm ex} \geq 2000$K) are expected to be formed
partially or fully inside the dust condensation radius. In this region, shocks may be important for the excitation
structure of molecules. We present the extracted values of all transitions.

Since the dust properties affect the excitation of the gas lines, our approach is to consistently fit gas and dust
emission. The dust emission was characterized by comparing our dust model to ISO observations of the dust excess \citep{Justtanont2004} and PACS and SPIRE
photometric measurements \citep{Groenewegen2011}.
A comparison of our model to the PACS image at 70 $\mu$m \citep{Cox2012} shows dust emission at distances beyond 20$\arcsec$ that is not reproduced by
our constant mass-loss rate dust model. We do not attempt to fit the 70 $\mu$m image but use it as evidence that the wind of W\,Hya will not be reproduced well by our model
beyond these distances.

\subsubsection{PACS spectra}
\label{sec:PACS_spectra}

The original observations were performed on 2010 August 25 (observation identifiers 1342203453 and 1342203454), but an anomaly onboard
the spacecraft resulted in the absence of data in the red channel. The observations were then rescheduled for 2011 January 14 (observation identifier
1342212604), band B2A, covering 51-73\,$\mu$m, and R1A, 102-146\,$\mu$m) and on 2011 July 9 (observation identifier 1342223808) band B2B,
covering 70-100\,$\mu$m and R1B, 140-200\,$\mu$m).

The data were reduced with Herschel Interactive Pipeline (HIPE) 10 and calibration set 45. The absolute flux calibration was performed via PACS internal calibration blocks and the
spectral shape derived through the PACS relative spectral response function. The data were rebinned with an oversampling factor of 2, which corresponds
to Nyquist sampling with respect to the resolution of the instrument. From a spectroscopic point of view, W\,Hya was considered a point source, because the transitions
observed by PACS are formed deep in the molecular wind, a region that is small compared to the beam size. The corresponding beam
correction was applied to the spectrum of the central spatial pixel.

Wavelength shifts greater than 0.02$\,\mu$m were observed between the two spectra obtained in band B2A.
In band B2B, the continua measured by both observations agree within 10\%. Still, the spectral lines as measured on the first observation date
were significantly weaker than those obtained on the later date. These two features indicate that there was a pointing error during the earlier observations. We have mitigated this by
only considering the data obtained on 2011 January 14 in our line-flux determinations. The effect of mispointing is marginal in the red, and no CO line was
detected shortwards of 70\,$\mu$m, so that the mispointing that affected the first observations has no impact on our results.

The CO lines were fitted using a Gaussian profile on top of a local, straight continuum. Whenever necessary, multiple profiles were fitted to a set of neighbouring
lines to account for blending and to improve the overall quality of the fit.
For the continuum fit we used two spectral segments, one on each side of the fitted group of lines.
These segments were taken to be four times the expected full width half maximum (FWHM) of a single line.

We considered an intrinsic error of 20\% on the measured line fluxes, owing to calibration uncertainties, on top of the error from the Gaussian fitting.

\subsubsection{SPIRE FTS spectra}
\label{sec:SPIRE_spectra}

The SPIRE FTS spectra were taken on 2010 January 9 (observation identifier 1342189116).
Seventeen scans were taken in high spectral resolution, sparse pointing
mode giving an on- source integration time of 2264 seconds. The native spectral resolution of the
FTS is 1.4 GHz (FWHM) with a sinc-function instrument line shape. The data were reduced using the
 HIPE SPIRE FTS pipeline version 11 (Fulton et al, in prep.) into a standard
spectrum of intensity versus frequency assuming W Hya to be a point source within the
SPIRE beam. To allow the CO lines to be simply fitted with a Gaussian profile, the data were
apodized to reduce ringing in the wings of the instrumental line function using the extended Norton-Beer
function 1.5 \citep{Naylor2007}. The individual CO lines were then fitted using the IDL
GAUSSFIT function. The fitted peak heights were converted to flux density using the measured
average line width of 2.18 GHz and the error on the fitted peak height taken as a measure of the
statistical uncertainty. The absolute flux calibration has been shown to be $\pm$\,6 \% for the SPIRE FTS
(Swinyard et al, in prep). We adopt a total uncertainty of 15 \% for the extracted line fluxes, to which we add
the errors of the individual Gaussian fits.

\begin{table}[!t!]
\caption{Observed $^{12}$CO and $^{13}$CO line fluxes for W\,Hya.}
\label{table:WHya_line_fluxes}
\scriptsize
\centering
\small{
\begin{tabular}{ | c  @{}c | r r r r @{$\pm$} l @{ }c | }     
\hline    

\multicolumn{2}{|c|}{Instrument} &  \multicolumn{1}{c@{}}{$J_{\rm up}$} & \multicolumn{1}{c}{$\phantom{a.}\nu_0$} & \multicolumn{1}{c@{}}{$E \ (K)$} & \multicolumn{2}{c}{Int. flux} & FWHM\\ 
\multicolumn{2}{|c|}{} & & [GHz] & [K] & \multicolumn{2}{c}{[$10^{-17}$ W/m$^2$]} & [arcsec] \\
\hline
\rule{0pt}{2ex}
& & \multicolumn{5}{c}{\bf $^{12}$CO} & \\
 &  SEST & 1 & 115.27 & 5.5 & \multicolumn{2}{c}{-\phantom{     }} & 43 \\

 & SMT & 2 & 230.54 & 16 & \multicolumn{2}{c}{-\phantom{   }} & 22 \\

 & APEX & 3 & 345.80 & 33 & \multicolumn{2}{c}{-\phantom{   }} & 17\\

 &   & 4 & 461.04 & 55 & \multicolumn{2}{c}{-\phantom{   }} & 13.5\\

& HIFI & 6 & 691.47 & 116 & $8.1 $ & $ 1.5$ & 29.5\\

& & 10 & 1151.98 & 304 & $15 $ & $ 3.3$ & 17.5 \\

& & 16 & 1841.34 & 752 & $21 $ & $ 3.0$ & 11.5\\

& SPIRE & 4 & 461.04 & 55 & $ 4.8 $ & $ 0.8 $ & 40.5\\

& & 5 & 576.27 & 83 & $ 6.2 $ & $ 1.0 $ & 32.6\\

& & 6 & 691.47 & 116 & $ 6.2 $ & $ 1.0 $ & 29.5\\

 & & 7 & 806.65 & 155 & $ 11 $ & $ 1.6 $ & 35.5\\

 & & 8 & 921.80 & 199 & $ 15 $ & $ 2.3 $ & 36.7\\

& & 9 & 1036.91 & 249 & $ 16 $ & $ 2.4 $ & 19.2\\

& & 10 & 1151.98 & 304 & $ 21^a $ & $ 4.0 $ & 17.5\\

& & 11 & 1267.01 & 365 & $ 19 $ & $ 1.3 $ & 17.6 \\

& & 12 & 1381.99 & 431 & $ 18 $ & $ 1.4 $ & 17.0 \\

& & 13 & 1496.92 & 504 & $ 17 $ & $ 1.4 $ & 16.8 \\

& PACS & 14 & 1611.79 & 581 & $ 20 $ & $ 5.0$ & 13\\

& & 15 & 1726.60 & 664 & $ 25 $ & $ 6.0$ & 12.5\\

& & 16 & 1841.34 & 752 & $ 39^b $ & $ 9.0$ & 11.5\\

& & 17 & 1956.02 & 846 & $ 27.5 $ & $ 5.5$ & 11\\

& & 18 & 2070.62 & 946 & $ 31 $ & $ 7.5$ & 10.5\\
                  
& & 19 & 2185.13 & 1050 & $ 25 $ & $ 6.0$ & 10\\
                  
& & 20 & 2299.57 & 1160 & $ 46.5^b $ & $ 13 $ & 9.8\\
                  
& & 21 & 2413.92 & 1276 & $ 21 $ & $ 6.0$ & 9.6\\
                  
& & 22 & 2528.17 & 1397 & $ 75^a $ & $ 20 $ & 9.4\\
                  
& & 23 & 2642.33 & 1524 & $ 140^a $ & $ 35 $ & 9.2\\
                  
& & 24 & 2756.39 & 1657 & $ 31 $ & $ 10 $ & 9\\
                  
& & 25 & 2870.34 & 1794 & \multicolumn{2}{c}{-\phantom{   }} & 9\\
                  
& & 26 & 2984.18 & 1937 & \multicolumn{2}{c}{-\phantom{   }} & 9\\
                  
& & 27 & 3097.91 & 2086 & $ 20 $ & $ 6.0 $ & 9\\
                  
& & 28 & 3211.52 & 2240 & $ 68 $ & $ 14 $ & 9\\
                  
& & 29 & 3325.00 & 2400 & $ 12 $ & $ 10 $ & 9\\
                  
& & 30 & 3438.36 & 2565 & $ 20 $ & $ 6.0 $ & 9\\

& ISO & 16 & 1841.34 & 752 & $33^c  $ & $ 10$ & 70 \\

& & 17 & 1956.02 & 846 & $22.6 $ & $ 10$ & 68\\

\hline
\rule{0pt}{2ex}

& & \multicolumn{5}{c}{\bf $^{13}$CO} & \\

& HIFI & 6 & 661.07 & 111 & $ 1.2 $ & $ 0.3 $ &  32 \\

& & 9 & 991.33 & 238 & $ 2.0 $ & $ 0.7 $ & 21 \\

& & 10 & 1101.35 & 291 & $2.8 $ & $ 0.9$ & 18 \\

\hline                  
\end{tabular}
}
\tablefoot{$a$ - Known blends with strong water
transitions excluded from the modeling procedure. b - flagged as blends from full-width half maximum of Gaussian fit. c - this
transition was flagged as a blend, since it was flagged as such during the PACS data reduction and ISO has a lower resolution
than PACS.}
\end{table}

\subsection{Observed line shapes}
\label{sec:lineshapes}

\begin{figure}
   \centering
   \includegraphics[width= 9cm]{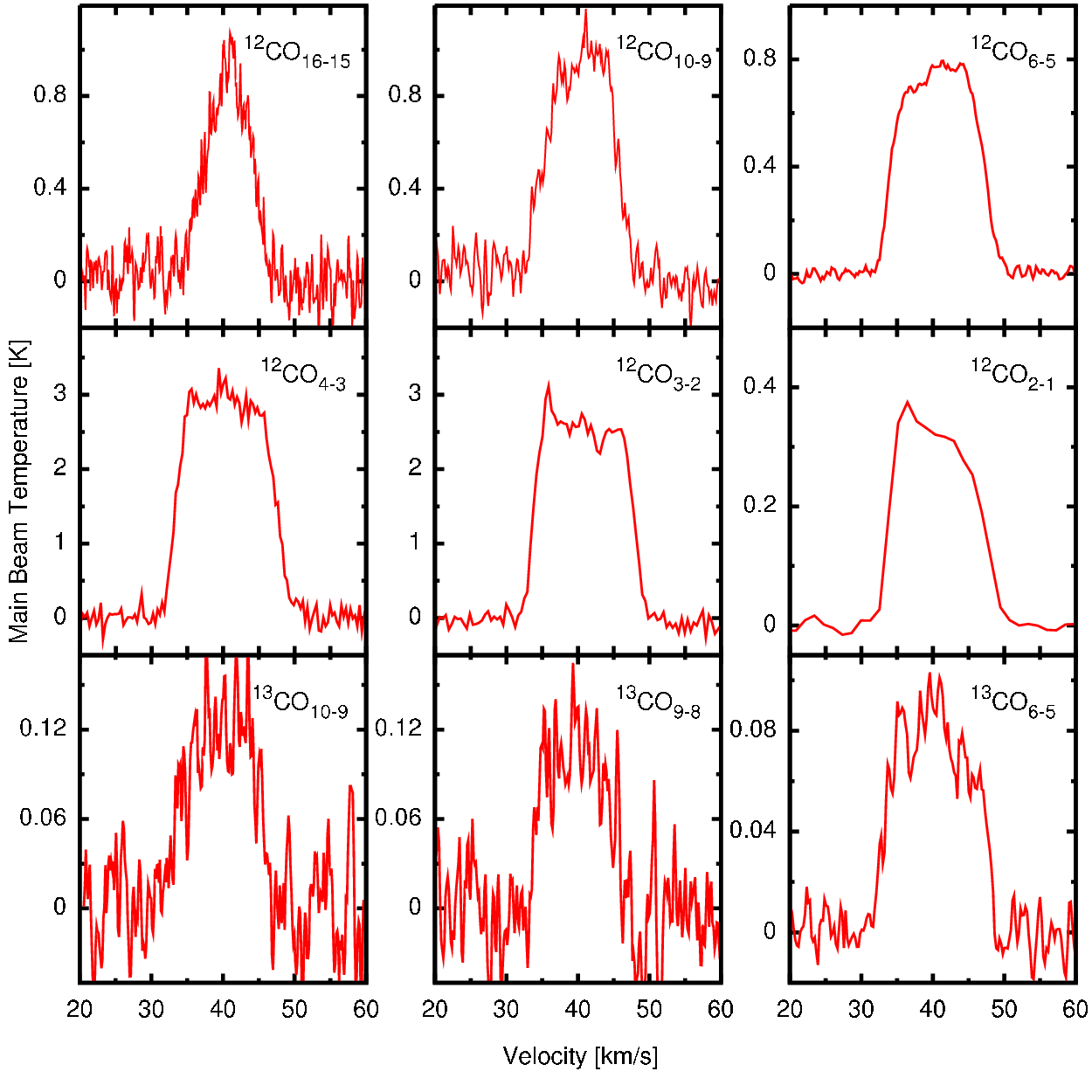}
      \caption{Observed line profiles of $^{12}$CO and $^{13}$CO. The data is from HIFI, APEX and SMT (see Table \ref{table:WHya_line_fluxes}).}
         \label{fig:symmetry}
   \end{figure}

The CO lines observed by HIFI, APEX, and SMT are shown in Fig. \ref{fig:symmetry}. A first inspection of the data
already unveils interesting properties of W\,Hya and indicates which region of the envelope is being probed by each transition.

The low and intermediate excitation transitions ($J_{\rm up} \le 6$) of $^{12}$CO show profiles characteristic of an optically thin wind. There seem to be no
indications of strong departure from spherical symmetry. These lines probe the outermost regions of the wind, where material
has reached the terminal expansion velocity.

The $J$=16-15 transition observed with HIFI shows a triangularly shaped profile characteristic of being formed in a region where the wind
has not yet reached the terminal velocity \citep{Bujarrabal1986}. This profile is, therefore, an important
tool for understanding the velocity profile of the accelerated material. The observations of transition $J$=16-15 were partially affected by
standing waves, and the vertical polarization could not be used. The horizontal polarization, in its turn, was not affected, and the line
shape and flux obtained from it are the ones used in this paper.

\subsection{Model assumptions}

To model $^{12}$CO and $^{13}$CO emission lines, we use the non-local thermodynamic equilibrium (non-LTE)
molecular excitation and radiative transfer code GASTRoNOoM \citep{Decin2006,Decin2010a}. We calculate the dust temperature
and emissivity as a function of radius with MCMax, a
Monte Carlo dust continuum radiative transfer code presented by \citet{Min2009}. The two codes are combined to provide a consistent physical description of
the gaseous and dusty components of the circumstellar envelope \citep{Lombaert2013}.

The central stellar emission was approximated by a $2500\ K$ black body with $5400\ L_{\odot}$, based
on the comparison with the ISO spectrum and adopting the value for W\,Hya's distance published by \cite{Knapp2003} (78 parsecs).
Based on these assumptions, our model has a stellar radius, R$_\star$, of 2.7 $\times 10^{13}$ cm.
Whenever we give values in units of stellar radii, we refer to this number.

We assumed a constant mass-loss rate and spherical symmetry. However, observations of W\,Hya's photosphere and the region
 immediately surrounding it, i.e. the molecular layer seen in IR bands and the radio photosphere, seem to be reproduced better by an
 ellipsoid than a sphere, and may even show a slow bipolar outflow or a rotating disk-like structure \citep{Vlemmings2011}.
As pointed out by \cite{Zhao-Geisler2011}, however, the position angles reported for the orientation of the ellipsoid
in the plane of the sky are sometimes contradictory, and therefore, no firm conclusions can be drawn on the exact value
of this observable. In an absolute sense, the departures from spherical symmetry reported are not large.
We conclude that using a spherical symmetrical model should not affect our findings in a significant way.

\subsubsection{MCMax and dust model}

MCMax calculates the dust temperature as a function of radius for the different species being considered.
\cite{Kama2009} implemented dust-sublimation calculations. In this way, grains are only present in regions
where they can exist with a temperature lower than their sublimation temperature. The inner radius of the
dust envelope is thus determined in a consistent way based on the dust species input. The dust temperature
and opacity as a function of radius output by MCMax are fed to the molecular line code GASTRoNOoM.

Our dust model is based on the work carried out by \cite{Justtanont2004},
who calculated dust extinction coefficients assuming spherical grains using Mie theory.
They took the grain size distribution to be the
same as found by \cite{Mathis1977} for interstellar grains. Following
\cite{Lombaert2013}, we instead represent the shape distribution of
the dust particles by a continuous distribution of ellipsoids
\citep[CDE,][]{Bohren1998, Min2003}. In the
CDE approximation, the mass-extinction coefficients are determined for
homogeneous particles with constant volume. The grain size, $a_{CDE}$, in this
context, is understood as the radius
of a sphere with an equivalent volume to the particle considered. This
approximation is valid in the limit where $a_\mathrm{CDE} \ll \lambda$.
We assume $a_\mathrm{CDE} = 0.1\ \mu$m.
A CDE model results in higher
extinction efficiencies relative to spherical particles and
more accurately reproduces the shape of the observed dust features \citep{Min2003}.

In our modelling, we include astronomical silicates
\citep{Justtanont1992}, amorphous aluminium oxides, and magnesium-iron oxides.
The optical constants for amorphous aluminium oxide and magnesium-iron oxide were
retrieved from the University of Jena database from the works of
\cite{Begemann1997} and \cite{Henning1995}, respectively.

\subsubsection{GASTRoNOoM}

The velocity profile of the expanding wind is parametrized as a $\beta$-type law,
\begin{equation}
\label{eq:beta-law}
\varv(r) = \varv_{\circ} + (\varv_{\infty} - \varv_{\circ}) \left(1 - \frac{r_{\circ}}{r} \right)^{\beta},
\end{equation}
where $\varv_{\circ}$ is the velocity at the dust-condensation radius $r_{\circ}$. We set $\varv_{\circ}$ equal to the local sound speed \citep{Decin2006}.
The flow accelerates up to a terminal velocity $\varv_{\infty}$.
The velocity profile in the wind of W\,Hya seems to indicate a slowly accelerating wind \citep{Szymczak1998}, implying a relatively high value of
$\beta.$ We adopt a standard value of $\beta$=1.5. We explore in Sect.
\ref{sec:vel_law} the impact of changing this parameter on the fit to the high-excitation line shapes.
The value of $r_{\circ}$ can be constrained from the MCMax
dust model, and is discussed in Sect. 4.2.4. The value of $\varv_{\infty}$ can be constrained from the width of low excitation CO lines, but depends
on the adopted value for the turbulent velocity, $\varv_{\rm turb}$. The velocity profile inside $r_{\circ}$ is also taken to be a $\beta$-type law but
with an exponent of 0.5, which is typical of an optically thin wind.

We parametrize the gas temperature structure of the envelope using a power law where
the exponent, $\epsilon$, is a free parameter,
\begin{equation}
\label{eq:temp}
T(r) = T_{\star}\ (R_{\star}/r)^{\epsilon}.
\end{equation}
This simple approach was motivated by uncertainties in the calculations of the cooling due
to water line emission in the envelope of AGB stars.
The final free parameter is related to the photo-dissociation of CO by an external ultraviolet radiation field.
We follow \cite{Mamon1988} who propose a radial behaviour for the CO abundance
\begin{equation}
\label{eq:r_1/2}
f_{\rm CO}(r) = f_{\rm CO}(r_\circ) \, {\rm exp}{\left[-{\rm ln}(2) \left(\frac{r}{{\rm r}_{\rm1/2}}\right)^{\alpha}\right]},
\end{equation}
where $f_{\rm CO} (r_{\circ})$ is the CO abundance at the base of the wind, and the exponent value $\alpha$ is set by the mass-loss rate. A value of 2.1 is suitable
for W Hya. The parameter $r_{1/2}$ represents the radius at which the CO abundance has decreased by half;
lower values of this parameter correspond to a more efficient
CO dissociation and therefore to a photodissociation zone that is closer to the star. The value found by \cite{Mamon1988} that suits
a star with parameters similar to W Hya's is referred to as $r_{1/2}^{\rm \,m}$.
We assume that both isotopologues have identical dissociation radii, since chemical fractionation is expected to compensate for the selective
dissociation \citep{Mamon1988}. The effects of changes in $r_{1/2}$ are discussed in Sect. \ref{subsub:just}.

We adopt a value of $2 \times 10^{-4}$ for the initial $^{12}$CO abundance number ratio relative to H, $f_{\rm CO}(r_\circ)$.
The effect of changes in this value on the derived $^{12}$CO / $^{13}$CO ratio is discussed in Sect. \ref{sec:disc}.
We consider the lowest 60 rotational levels
of both the ground and first-excited vibrational levels of the $^{12}$CO and $^{13}$CO molecules. The adopted radiative and collisional coefficients
are identical to those adopted by \cite{Decin2010a}. The code takes the beam shape of the telescope into
account, allowing for a direct comparison of predicted and observed line shapes.

\section{The isotopic ratio from observed pure rotational transitions of $^{12}$CO and $^{13}$CO}
\label{sec:radiation_field}

For the range of gas temperatures for which CO molecules are formed and observed in a low mass-loss
rate AGB star, such as W\,Hya, we can assume that the isotopologic $^{12}$CO/$^{13}$CO ratio reflects the isotopic $^{12}$C/$^{13}$C ratio directly.
Therefore, the measured ratio of integrated line fluxes of corresponding pure rotational transitions of $^{12}$CO and
$^{13}$CO are good probes of the intrinsic $^{12}$C\,/\,$^{13}$C abundance ratio.
However, owing to the different abundances of these two molecules, the two excitation structures are likely to be significantly different.
This means that retrieving isotopologic ratios from observed flux ratios of pure rotational lines is not trivial.
Understanding how different these excitation structures can be and how they affect the observed line intensities is a powerful tool
when trying to retrieve the intrinsic isotopic ratio. In this section, we explore this mechanism and discuss important observables that
may help constrain the $^{12}$CO/$^{13}$CO ratio.

As discussed by \cite{Morris1980}, there are two sharply defined excitation regimes for
CO molecules: when they are collisionally excited or when excitation is dominated by the near infrared ($\approx 4.6\ \mu$m) radiation field.
$^{13}$CO responds more strongly to excitation governed by the infrared radiation field, since this is the less abundant isotopologue, and the medium is more transparent
at the wavelengths of the relevant ro-vibrational transitions.
We find that in a low mass-loss rate star, such as W\,Hya where the stellar radiation field is the main source of infrared photons, $^{13}$CO
molecules are more efficiently excited by direct stellar radiation to higher rotational levels further away from the star than $^{12}$CO.
\cite{Schoier2000} discuss that this effect is important even for high mass-loss rate carbon stars.

For low mass-loss rate AGB stars, similar to W\,Hya, the observed $^{12}$CO and $^{13}$CO rotational transitions are mostly optically thin, with the highest tangential optical depths reached being
around one. Therefore, the total emission of a given observed
transition is roughly proportional to the total number of molecules in the corresponding upper rotational level.

\begin{figure}
\centering
\includegraphics[width= 9cm]{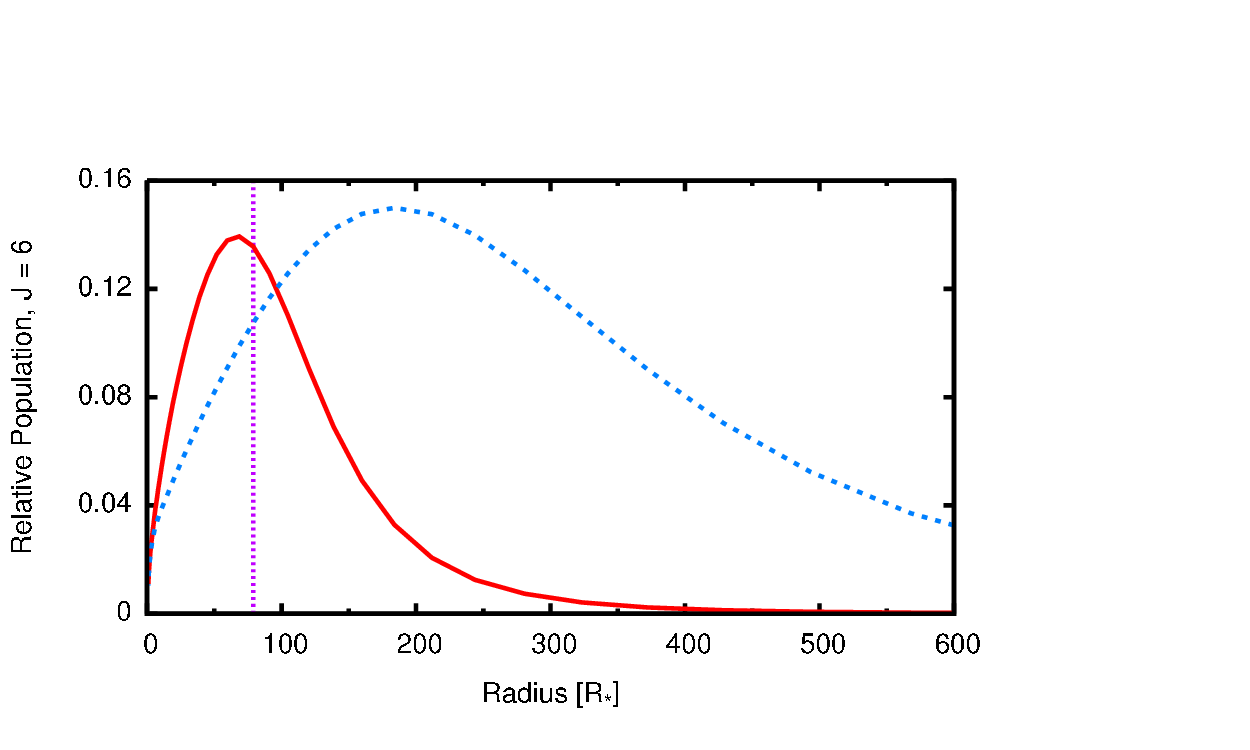}
\caption{Fractional population of level $J$=6 from a model for a low mass-loss rate AGB star for $^{12}$CO (full red line) and $^{13}$CO (blue dashed line),
with an isotopic ratio $^{12}$CO/$^{13}$CO = 20. The vertical purple dashed line marks the point where critical density is reached for $^{12}$CO, calculated
using the collisional coefficients between H$_2$ and CO for a 300 K gas. $^{13}$CO
reaches the critical density at a 5\% greater distance.}
\label{fig:J_6_12CO_13CO}
\end{figure}

\begin{figure}
   \centering
   \includegraphics[width= 9cm]{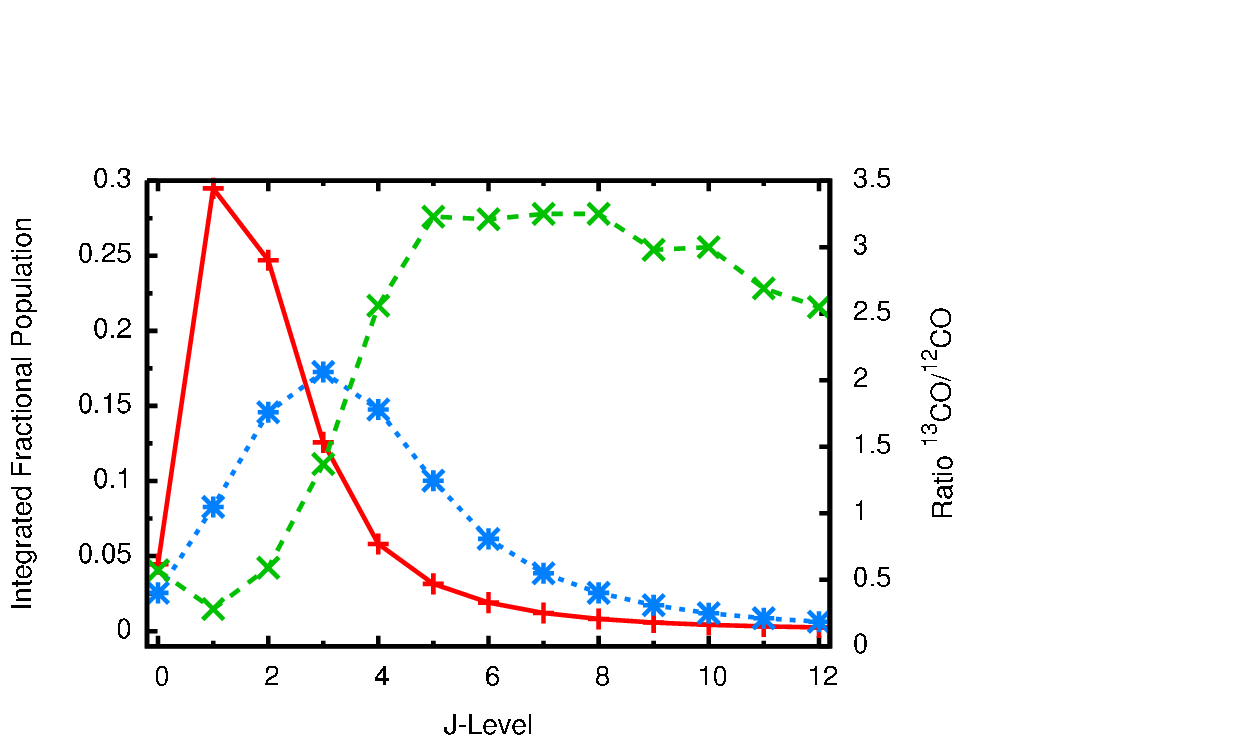}
      \caption{Integrated relative $^{12}$CO and $^{13}$CO populations in a low mass-loss rate model envelope with an isotopic ratio $^{12}$CO/$^{13}$CO = 20. $^{12}$CO is represented by the red solid line;
      $^{13}$CO, by the blue short-dashed line; and the ratio of the two, shown on the right-hand axis by the green long-dashed line.}
         \label{fig:Integrated_12CO_13CO}
   \end{figure}

In Figure \ref{fig:J_6_12CO_13CO}, we show the relative population of level
$J$=6 for both $^{12}$CO and $^{13}$CO calculated in a model
for a low mass-loss rate star. We define the relative population of a given rotational level
$J$ as $f_J(r) = \frac{{\rm n}_J({\rm r})}{{\rm n}_{\rm CO}({\rm r})}$, where ${\rm n}_J({\rm r})$ and
${\rm n}_{\rm CO}({\rm r})$ are, respectively, the number densities at a distance r of CO molecules excited to level $J$
and of all CO molecules.
As can be seen, the same level for the two molecules probes quite different regions of the envelope, and the relative
amount of $^{13}$CO molecules in level $J$=6 is higher than for $^{12}$CO. In Fig. \ref{fig:Integrated_12CO_13CO}, we show that this effect
influences the rotational levels of these two molecules different by plotting the integrated relative population,
$\frac{\int_{\rm R_\star}^{\rm R_{\infty}}{\rm n}_J({\rm r})dr}{\int_{\rm R_\star}^{\rm R_{\infty}}{\rm n}_{\rm CO}({\rm r})dr}$,
 of up to $J$=12. Therefore, observations of different rotational lines of both isotopologues
will result in very different line ratios. Owing to the more efficient excitation of $^{13}$CO to higher rotational levels further away from the star, the low-$J$ lines are expected to show a higher $^{12}$CO/$^{13}$CO
line flux ratio value than the intrinsic isotopic ratio, and intermediate-$J$ lines are expected to show a lower value than the intrinsic isotopic ratio.

The point where the envelope reaches the critical density, below which collisional de-excitation is less probable than spontaneous de-excitation, is also shown in Fig. \ref{fig:J_6_12CO_13CO}.
Parameters such as the total H$_2$ mass-loss rate, which indirectly set in our models by the $^{12}$CO abundance, and the stellar luminosity control the location of
the point where the critical density is reached. The infrared stellar radiation field will have an impact on the excitation structure of the two molecules, mainly of
$^{13}$CO, especially when the density is below critical. Therefore, uncertainties on the $^{12}$CO abundance, the stellar luminosity, and the infrared stellar
radiation field will very likely lead to uncertainties on the determined isotopic ratio.
Interferometric observations might be an important tool for determining the $^{12}$CO/$^{13}$CO ratio more precisely, since those would show the region where
emission originates for a given level of $^{12}$CO and $^{13}$CO, hence constraining the $^{12}$CO abundance and the importance of selective excitation of $^{13}$CO.

\section{Model for W\,Hya}
\label{sec:CO_WHya}

\subsection{Dust model}
\label{sec:dust_model}

The model for the $^{12}$CO and $^{13}$CO emission from W\,Hya considers a dust component based on the parameters found by
\cite{Justtanont2004}, but fine-tuned to match our different approximation for calculating the
extinction coefficient. This yielded slightly different abundances for each species and a smaller total dust mass-loss rate.
We briefly discuss the dust model, stressing the points that are relevant for the CO emission analysis.

\begin{figure}[h]
   \centering
   \includegraphics[width= 8.5cm]{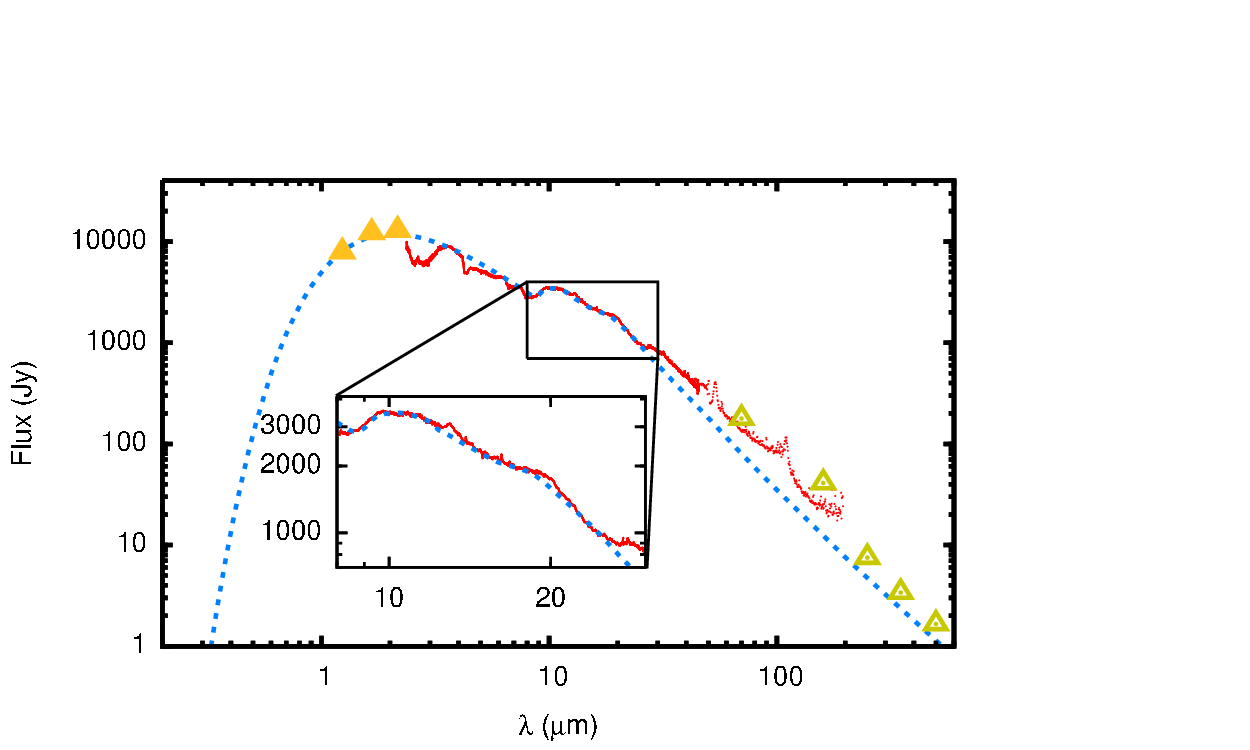}
      \caption{Dust model (blue dashed line) compared to the ISO (red) and PACS and SPIRE (open-yellow triangles) photometric observations
      and to near-IR photometric observations (filled-yellow triangles).
      The fit to the silicate and amorphous aluminium oxide spectral features is shown in detail. The feature at 13 $\mu$m is not taken into account in our fit.}
         \label{fig:fit_ISO}
   \end{figure}

A total dust mass-loss rate of $2.8 \times 10^{-10}$ M$_{\odot}$ yr$^{-1}$ is needed to reproduce the SED with our dust model.
Our best fit model contains 58 \% astronomical silicates,
34 \% amorphous aluminium oxide (Al$_2$O$_3$), and 8 \% magnesium-iron oxide (MgFeO).
We prioritized fitting the region between 8 and 30 $\mu$m of the IR spectrum (see Fig. \ref{fig:fit_ISO}).
However, we did not attempt to fit the 13 $\mu$m feature seen in the ISO spectrum of W\,Hya, since its origin is still a matter of debate.
Candidate minerals that might account for this feature are crystalline aluminium oxides, either crystalline corundum ($\alpha$-Al$_2$O$_3$) or
spinel (MgAl$_2$O$_4$). One of these two species, at an abundance of less than 5\%, might account for the observed 13 $\mu$m feature \citep[see][and references therein]{Posch1999,Zeidler2013}.
We estimate an uncertainty on each of the derived dust abundances of 30 \%.

\begin{figure}[h]
   \centering
   \includegraphics[width= 8.5cm]{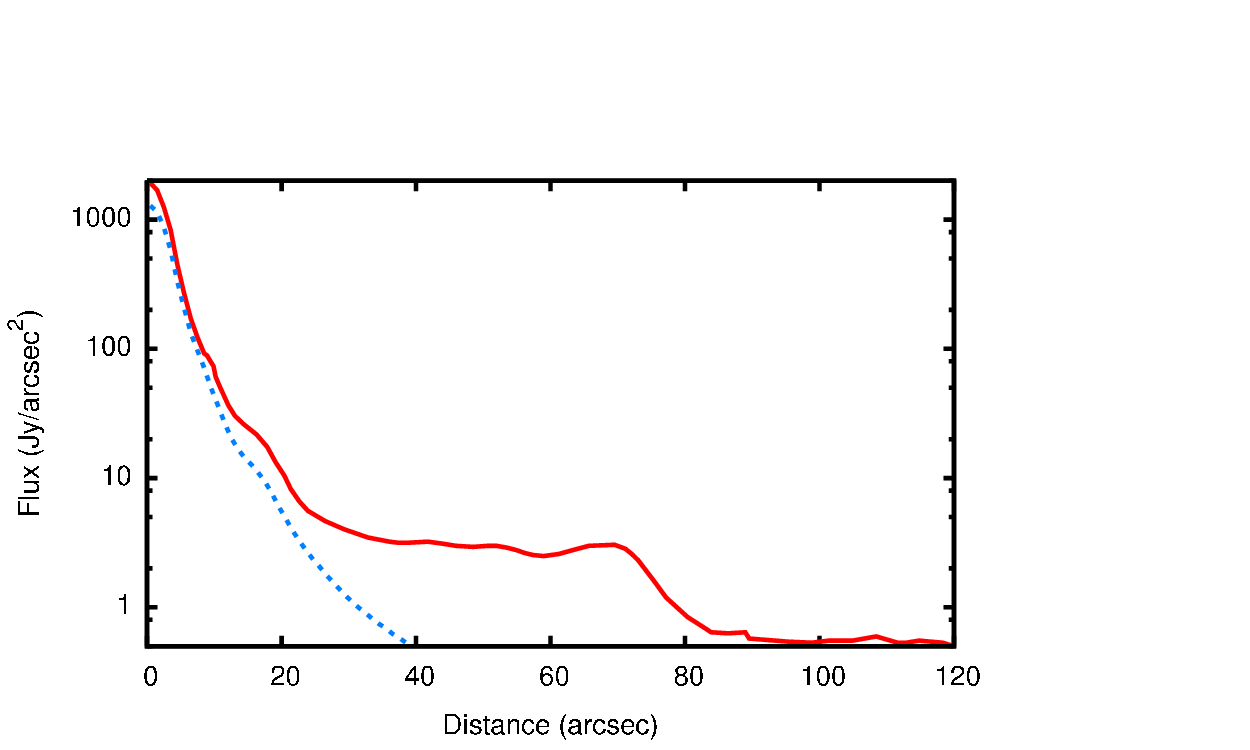}
      \caption{Dust model (dashed blue) compared to the $70\ \mu$m PACS image \citep[red solid line;][]{Cox2012}.}
         \label{fig:fit_PACS}
   \end{figure}

The dust condensation radius found from the sublimation calculations is around $2.5\ R_{\star}$
for amorphous aluminium oxide (the species expected to condense first). This value is very similar to the $2\ R_{\star}$ inner radius of the shell of large transparent grains recently observed
by \cite{Norris2012}. The condensation radius found for the astronomical silicates is at $5\ R_{\star}$, approximately a factor of 5 lower than
inferred by \cite{Zhao-Geisler2011} from interferometric observations. However, these authors calculated an average of several observations taken
at different epochs, and this mean is dominated by the data taken during mid-IR maximum. The ISO spectrum we used was obtained shortly after the visual minimum phase.
This may explain part of the discrepancy.

The dust model is compared to the ISO spectrum in Fig. \ref{fig:fit_ISO} and to the recently published PACS $70\ \mu$m maps in Fig. \ref{fig:fit_PACS}.
In the near infrared, W\,Hya's spectrum is dominated by molecular absorption bands, which are not included in our models.

That the assumption of constant mass loss fits the $70\ \mu$m PACS maps up to about 20 arcseconds (corresponding to 800 R$_\star$ for our adopted stellar parameters and distance)
does not necessarily imply that the envelope is explained well by a constant dust mass-loss rate within that radius, since the contribution from the
complex point-spread function is still important up to these distances. The good fit to the ISO spectrum between 10 $\mu$m and 30 $\mu$m, however, shows that the
inner dust envelope is reproduced well by a constant mass-loss rate model.
Interestingly, the $70\ \mu$m PACS image shows additional dust emission beyond 20 arcseconds from the star, which is not expected on the basis of
our model. Also, our model under-predicts the ISO spectrum from 30 $\mu$m onwards. It is possible that these two discrepancies between our model and the observations could be explained by
the extra material around the star seen in the PACS images.

Furthermore, the point where the dust map shows extra emission, 800 R$_{\star}$, coincides with what is roughly expected
for the CO dissociation radius from the predictions of \cite{Mamon1988} for a gas mass-loss rate of $1.5\times10^{-7}$ M$_\odot$yr$^{-1}$.
Since much of the extra dust emission comes from outside of that radius, and we do not expect the extra supply of far-infrared photons produced
by this dust to affect the CO excitation in a significant way, we will address the problem of the dust mass-loss history in a separate study. For the work carried out in this paper,
it is important to stress that there seems to be an extra amount of material around the region where CO is typically expected to dissociate for a star such as W\,Hya.
The origin of this extra material has not yet been identified and our model does not account for it.

\subsection{Model for CO emission}

Using GASTRoNOoM, we calculated a grid of models around the parameter values that are most often reported in the literature for W\,Hya.
We used a two-step approach. First, the modelled line fluxes were
compared to the SPIRE, PACS, HIFI, ISO, and ground-based observations using a reduced $\chi^{2}$ fit approach. Then, the line shapes were compared by eye to the normalized lines
observed by HIFI and ground-based telescopes. Once we had found the region of parameter space with the lower reduced $\chi^{2}$ values and good line shapes fit, we refined the grid
in this region in order to obtain our best grid model. The model parameters that are explored in our grid are presented in Table \ref{tab:grid_par}, together with the
best fit values found by us.

\begin{table}[h]
\caption{Range of model parameters used for grid calculations.}
\centering
\label{tab:grid_par}
\begin{tabular}{ c | c | c }     
\hline
Parameter & Grid values & Best fit \\
\hline
\hline
\rule{0pt}{2ex}
$dM/dt$ [$10^{-7}$ M$_{\odot}$yr$^{-1}$]& 1.0; 1.5; 2.0; 2.5& $1.5 \pm 0.5$\\
v$_{\infty}$ [km/s] & 6.5; 7.0; 7.5; 8.0 & $7.5 \pm 0.5$\\
$\epsilon$ & 0.5; 0.6; 0.7; 0.8 & $0.65 \pm 0.05$\\
v$_{\rm turb}$ [km/s] & 0.5; 1.0; 1.5; 2.0 & $1.4 \pm 1.0$ \\
$f_{\rm CO}$ & $2 \times 10^{-4}$ &$ 2 \times 10^{-4}$ \\
r$_{1/2}$ & r$_{1/2}^{\rm \,m}$ & r$_{1/2}^{\rm \,m}$\\
T$_{\star}$ [K] & 2500 & 2500 \\
\hline
\end{tabular}
\end{table}

\subsubsection{Result from grid calculation}

The best grid models have a terminal velocity of 7.5 km/s, $\epsilon=0.6$ or 0.7,
and a mass-loss rate of $1.5 \times 10^{-7}$ M$_{\odot}$yr$^{-1}$.
The turbulent velocity may be determined from the steepness of the line wings, but
could not be well constrained. Models with values of 1.0 km/s or higher all produce reasonable fits.

\begin{figure}[h]
   \centering
   \includegraphics[width= 8.5cm]{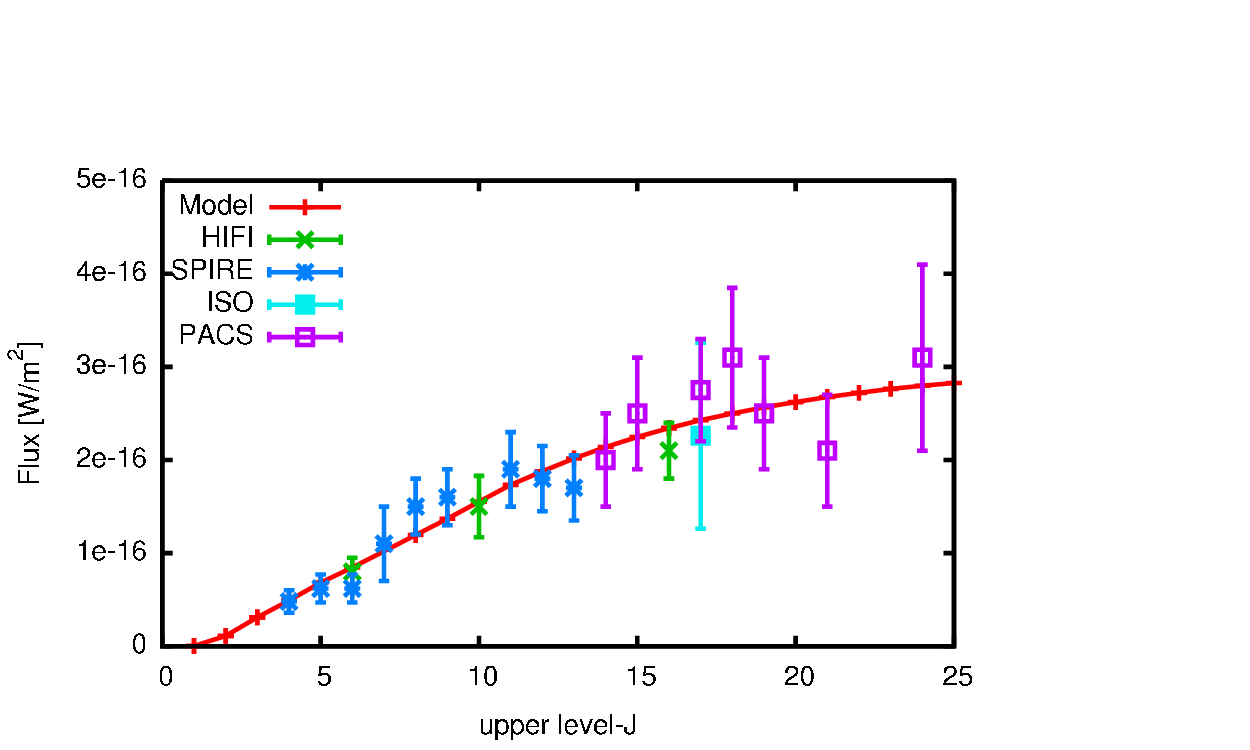}
      \caption{Best grid model for the $^{12}$CO observed integrated line fluxes, when considering the dissociation radius as given by \cite{Mamon1988}.
        Transition $^{12}$CO $J$=16-15 observed by PACS and ISO is not used due to line blending (see Table \ref{table:WHya_line_fluxes}).}
         \label{fig:mamon_fit_flux}
   \end{figure}

\begin{figure}[h]
   \centering
   \includegraphics[width= 8.5cm]{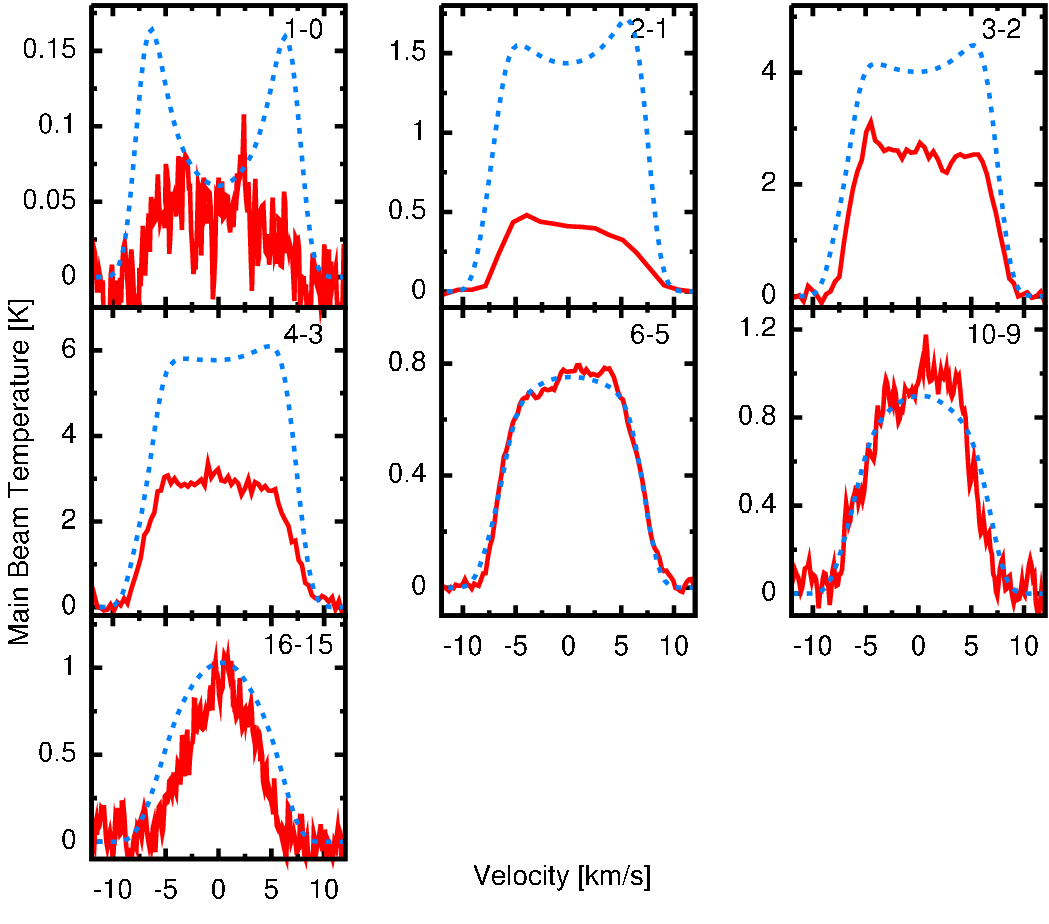}
      \caption{ Best grid model (in dashed blue) for the observed $^{12}$CO  line shapes (in solid red), when considering the photodissociation radius as given by
      \cite{Mamon1988} compared to the observed line shapes. Observations were carried out by HIFI, APEX, SMT, and SEST (see Table \ref{table:WHya_line_fluxes}).}
         \label{fig:mamon_fit_shape}
   \end{figure}

\begin{figure*}
   \centering
   \includegraphics[width= 17.5cm]{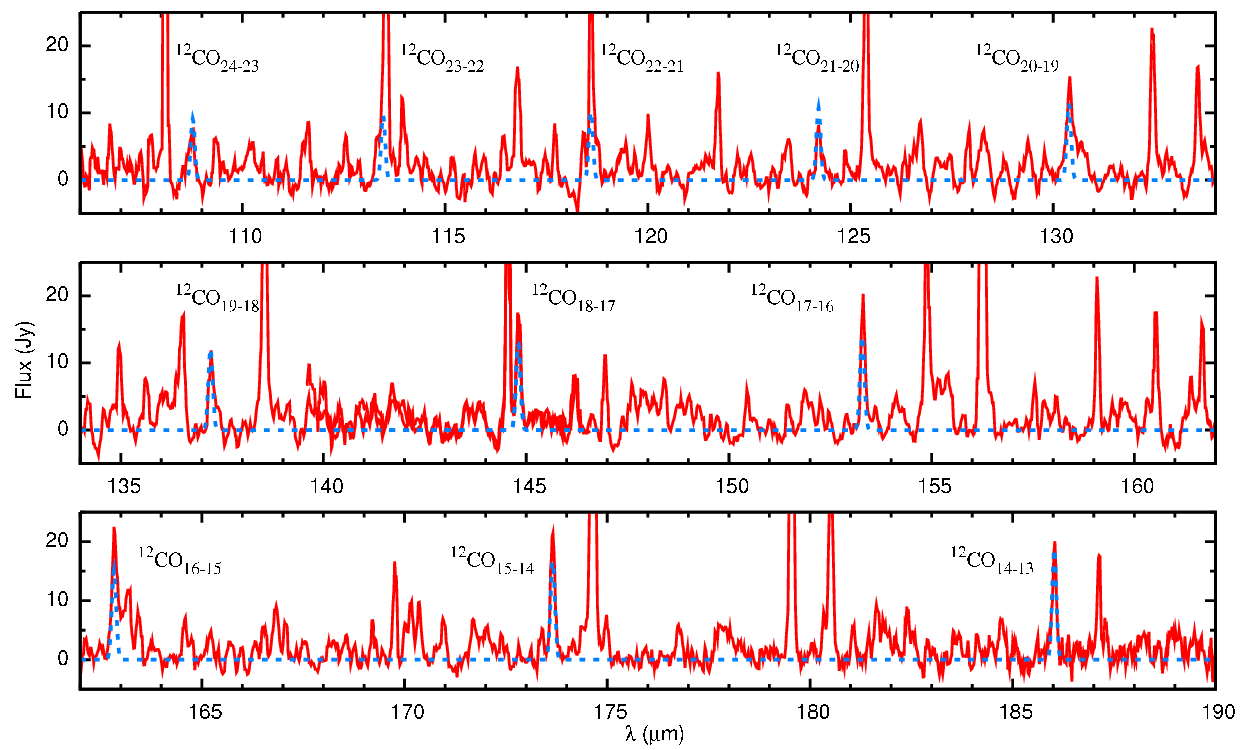}
      \caption{ Best grid model (in dashed blue) for the $^{12}$CO compared to the PACS spectrum (in solid red).
      Transition $J$=23-22 of $^{12}$CO is blended with the water transition $J_{\rm K_a,K_c}=4_{1,4}$-3$_{0,3}$ at 113.54 $\mu$m line, and transition $J$=22-21 is probably blended with
      both water transitions $J_{\rm K_a,K_c}=9_{3,  7}$-8$_{4,  4}$ and $\nu_2$=1, $J_{\rm K_a,K_c}=4_{3,  2}$-5$_{0,5}$ at 118.41 $\mu$m and 118.98 $\mu$m, respectively. }
         \label{fig:mamon_fit_pacs}
   \end{figure*}

Figures \ref{fig:mamon_fit_flux}, \ref{fig:mamon_fit_shape}, and \ref{fig:mamon_fit_pacs}
show the comparison of the best grid prediction with the observed integrated line fluxes, line shapes, and the PACS spectrum, respectively.

The overall fit of our best grid model to the line fluxes observed by SPIRE, PACS, HIFI, and ISO is quite good, and it does reproduce the broad trend of W\,Hya's
$^{12}$CO line emission. However, when we compare our model with the line shapes observed by HIFI and the ground-based measurements, we notice two things. First, as can be seen from Fig.
\ref{fig:mamon_fit_shape}, we over-predict the fluxes of the low-excitation $^{12}$CO lines, $J$=4-3, $J$=3-2, $J$=2-1, and $J$=1-0. Moreover, we predict the profiles to be double-peaked, while
the observations do not show that. Since double-peaked line profiles are characteristic of an emission region that is spatially resolved by the beams of the telescopes,
W\,Hya's $^{12}$CO envelope seems to be smaller than what is predicted by our models.
We discuss the problem of fitting the low-excitation lines in Sect. \ref{subsub:just}.
Second, for the highest excitation line for which we have an observed line shape ($^{12}$CO $J$=16-15), our model not only over-predicts the flux
observed by HIFI but also fails to match the width of the line profile. The observed triangular shape is characteristic of lines
formed in the wind acceleration region \citep[e.g.][]{Bujarrabal1986}. This indicates that either the wind acceleration is slower (equivalent to a higher $\beta$) or that the onset of the acceleration is more distant than the
2.5 R$_\star$ that is assumed in our grid and that corresponds to the condensation radius for amorphous Al$_{2}$O$_{3}$. This finding also implies that the
velocity law may be more complex than the $\beta$-law Eq.~\ref{eq:beta-law}. We return to this in Sect. \ref{sec:vel_law}.

\subsubsection{$^{13}$CO abundance}
\label{sec:13COMamon}

Using our best grid model parameters, we calculated a small second grid of models,
varying only the $^{12}$CO/$^{13}$CO ratio and assuming an isotopologic ratio that is constant throughout the envelope.

\begin{figure}[h!]
\centering
\includegraphics[width=8cm]{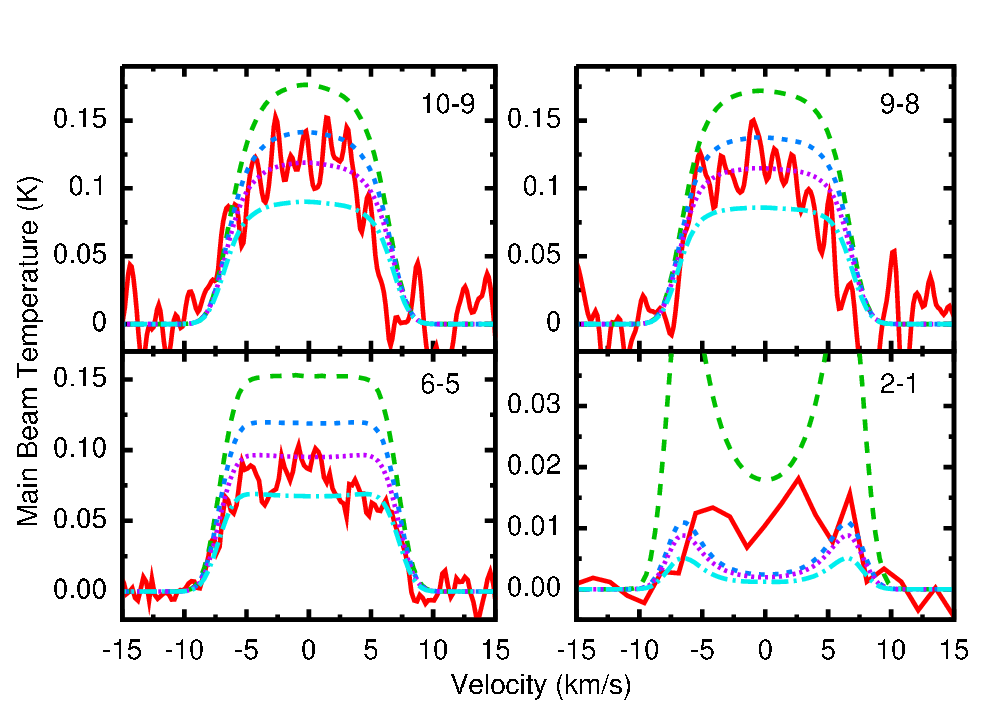}
\caption{Model calculations for different $^{12}$CO / $^{13}$CO ratios compared to the observed $^{13}$CO lines adopting the dissociation radius
by Mamom et al. (1988). The solid red line represents the observations. The long-dashed green, short-dashed blue, dotted purple, and
dot-dashed light blue lines are for intrinsic isotope ratios 10, 15, 20, and 30, respectively. Transitions $J$=10-9 and 9-8 have
been smoothed for a better visualization.}
\label{fig:13CO_10-9}
\end{figure}

The results are presented in Fig. \ref{fig:13CO_10-9}. An isotopologic ratio of 20 fits the $J$=10-9 and $J$=9-8 transitions but
somewhat over-predicts the $J$=6-5 one. Transition $J$=2-1 is predicted to be spatially resolved by SMT, though that is not supported by the observations.

\begin{figure}[h!]
\centering
\includegraphics[width=8cm]{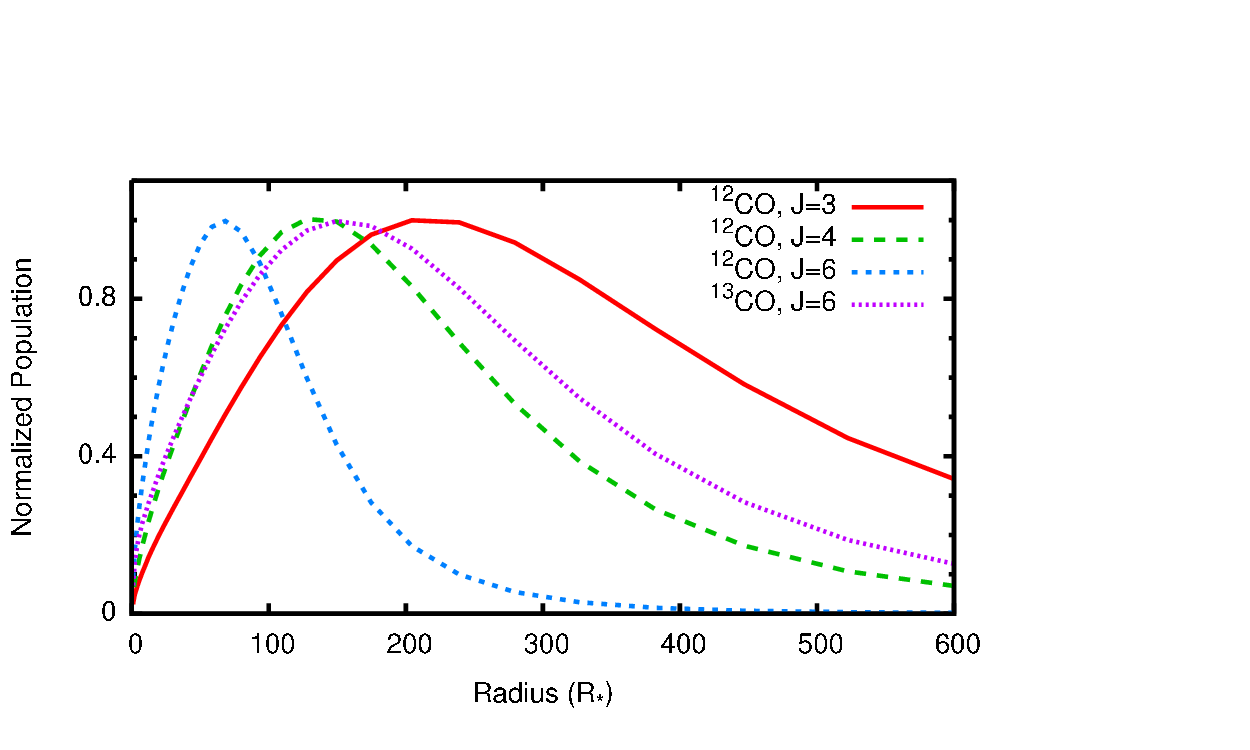}
\caption{Relative population distribution of different levels of $^{12}$CO and $^{13}$CO for our best grid model.}
\label{fig:12CO_13CO_levels}
\end{figure}

As discussed in Sect. \ref{sec:radiation_field}, our models show that these two molecules have very different excitation structures throughout the envelope.
In Fig. \ref{fig:12CO_13CO_levels} we compare the excitation region of the upper levels of the $^{12}$CO $J$=3-2, $^{12}$CO $J$=4-3,
$^{12}$CO $J$=6-5, and $^{13}$CO $J$=6-5 transitions for our best grid model.
This comparison shows that the $J$=6 level of $^{13}$CO has a similar excitation region to the $J$=4 level for $^{12}$CO. Our probes of the
outer wind are the line shapes and strengths of the $^{12}$CO $J$=4-3 and lower excitation lines and the line strength of the $^{13}$CO $J$=6-5 line.
They all point in the direction of fewer $^{12}$CO and $^{13}$CO molecules in these levels in the outer wind than our model predicts.

\subsubsection{The $\varv_{\rm LSR}$ of W\,Hya}

A value of 40.7 km/s is commonly found for the local standard of rest velocity, $\varv_{\rm LSR}$, of W\,Hya \citep[e.g.][]{Justtanont2005}.
However, there is still some uncertainty on this value, as determining it through different methods leads
to slightly different results. For example, \cite{Uttenthaler2011} find a value of 40.5 km/s, while \cite{Etoka2001} suggest 40.6 $\pm$ 0.4.

The value of $\varv_{\rm LSR}$ that best fits our data is 40.4 km/s. By shifting the modeled lines by this amount, the model fits the lines of $^{12}$CO and transition $J$=6-5 of $^{13}$CO well.
For the $J$=10-9 and $J$=9-8
transitions of $^{13}$CO, a $\varv_{\rm LSR}$ of 39.6 km/s seems more appropriate. These two lines are also narrower than our model predicts.
This difference points to an asymmetry in the wind expansion, the blue shifted-wing being accelerated faster than the red-wing one.
A direction-dependent acceleration from the two higher excitation lines of $^{13}$CO is not, however, conclusive at this point.

\subsubsection{Fitting the high-excitation lines: velocity profile}
\label{sec:vel_law}

A value of 1.5 for the exponent of the $\beta$-type velocity profile over-predicts the width of
$^{12}$CO $J$=16-15 observed by HIFI.
To reproduce the narrow line profile observed for this high excitation transition, we have to increase $\beta$ to about 5.0. This corresponds
to a very slowly accelerating wind. Increasing $\beta$ affects not only the line shapes but also the line fluxes. To maintain the fit to the observed fluxes,
we have to decrease the mass-loss rate by about 10\% to 1.3 $\times 10^{-7}$ M$_{\odot}$ yr$^{-1}$. With this change, the low-excitation lines also become roughly 10\%
weaker. This helps, but does not solve the problem of fitting these lines.

\begin{figure}[h!]
\centering
 \includegraphics[width=8.5cm]{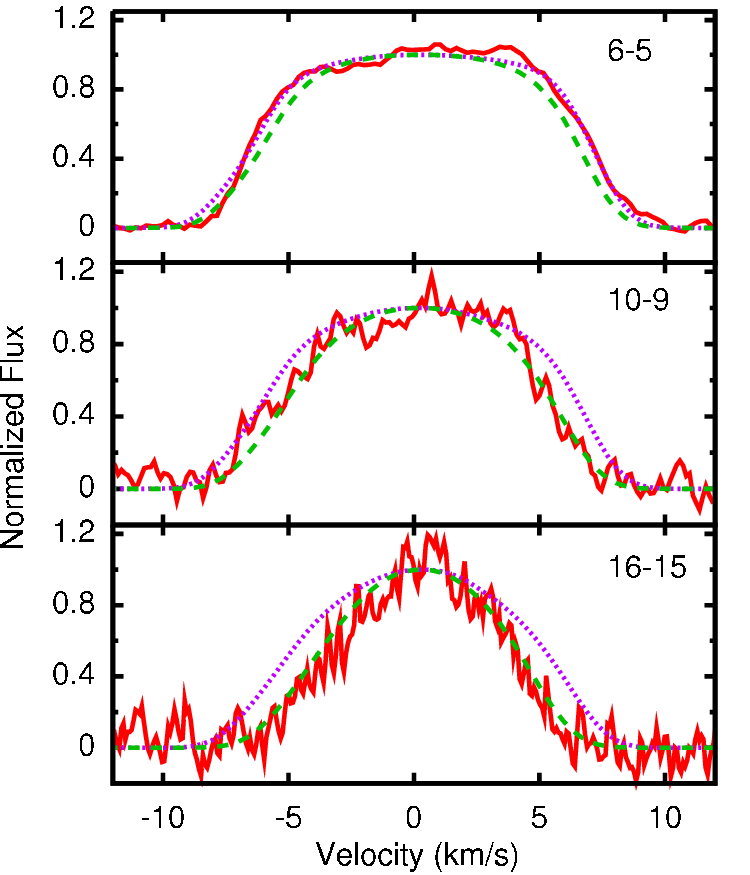}
  \caption{Normalized predicted line shapes for different values of the velocity law exponent compared to the normalized observed line shapes of the $^{12}$CO transitions
  observed by HIFI. The observations are shown by the solid red line,
   the dashed green line represents a model with $\beta$ = 5.0, and the dotted purple line a model with $\beta$ = 1.5.}
  \label{fig:beta_shapes}
 \end{figure}

The line shapes of transitions $J$=10-9 and $J$=6-5 of $^{12}$CO and $J$=10-9 and $J$=9-8 of $^{13}$CO (see Figs.
\ref{fig:beta_shapes} and  \ref{fig:13CO_smaller_radius}) are strongly affected by these changes.
In Fig. \ref{fig:beta_shapes} we show the comparison of our best grid model with the model with $\beta$ = 5.0 and lower mass-loss rate. As can
be seen, the line shape of transition $^{12}$CO $J$=16-15 is much better fitted by the higher value of $\beta$. For transition $^{12}$CO $J$=10-9, that is also the case for the red wing
but not for the blue wing. For transition $^{12}$CO $J$=6-5, the model with $\beta = 5.0$ predicts too narrow a line, while the model with $\beta = 1.5$ matches
the red wing but slightly over-predicts the emission seen in the blue wing.

Another possibility for decreasing the width of the high excitation lines is to set the starting point of the acceleration of the
wind, r$_{\circ}$ in Eq. \ref{eq:beta-law}, further out.
For those of our models discussed so far, the starting point of the acceleration is set at the condensation radius
of the first dust species to form (amorphous Al$_2$O$_3$), which is at 2.5 $R_{\star}$.
We need to set r$_{\circ}$ to 8 $R_{\star}$  to get a good fit to the line profile of transition $^{12}$CO $J$=16-15. As when $\beta$ is changed, we need to decrease
the mass-loss rate by approximately 10\% to compensate for the higher density in the inner wind .
 
Increasing the value of r$_{\circ}$ has a similar impact to increasing the value of $\beta$ and, therefore, models with different combinations of these parameters will be able to reproduce
the observed line shapes. That the wind has a slower start than expected, either because the gas acceleration starts further out or   a very gradual acceleration
is consistent with previous interferometric measurements of SiO, H$_2$O, and OH maser emission \citep{Szymczak1998} and of SiO pure rotational emission \citep{Lucas1992}.

From the final values found for the dust ($2.8 \times10^{-10}$ M$_{\odot}$ yr$^{-1}$) and gas (1.3 $\times 10^{-7}$ M$_{\odot}$ yr$^{-1}$) mass-loss rate, we derive a value for the dust-to-gas ratio of
$2 \times 10^{-3}$. Considering an uncertainty of 50 \% for both the derived dust and the gas mass-loss rates, we estimate the uncertainty on this value to be approximately 70\%.

\subsubsection{Fitting the low-excitation lines}
\label{subsub:just}

We now investigate why, within the parameter space region spanned by our grid calculation,
we cannot simultaneously fit the line fluxes of the intermediate and high excitation lines ($J_{\rm up} \ge 6$) and the line
fluxes and shapes of the low excitation lines ($J_{\rm up} \le 4$).

One possible cause could be a variable mass loss. In this case, the low excitation transitions
would be tracing gas that was ejected when the mass-loss rate was lower than what it is presently, which is traced by the
higher excitation lines.
This would be in strong contrast, however, to what we expect from the spatial distribution of the dust, since the PACS 70 $\mu$m image shows that
there is more dust emission at distances greater than the 20 arcseconds ($\approx$ 800 R$_\star$) that our dust model can account for. To reconcile this, we would
have to invoke not only a change in mass-loss rate but also quite an arbitrary and difficult-to-justify change in the dust-to-gas ratio.
Therefore, we argue that a mass-loss rate discontinuity is not a likely explanation for the problem of fitting the low
excitation lines in our model of W\,Hya.

Another possibility would be to consider a broken temperature law with a different exponent for the outer region where the low-excitation
lines are formed. We can either increase the exponent given in Eq. \ref{eq:temp} to obtain lower temperatures in the outer
regions or decrease the exponent to have higher temperatures in the outer regions. Changing the gas temperature does not necessarily
have a big impact on the $^{12}$CO level populations because the radiation field also plays an important role in the excitation.

We calculated models with both higher (0.9) and lower (0.4) values of $\epsilon$ for the outer regions assuming two different breaking points
for the temperature power law, either at 80 or 150 $R_{\star}$. The breaking points were chosen based on the excitation region of the low-excitation
lines, since at 150 $R_{\star}$ the population of level $J$=4 of $^{12}$CO peaks. The models with a breaking point at 80 $R_{\star}$ represents an intermediate step between introducing
a breakpoint that will only affect the low-excitation lines and changing the exponent of the temperature law in the entire wind. We find that the
strength of the low-excitation lines relative to each other changes but that the overall fit does not improve, since most of the lines are still over-predicted.
This suggests that we need a lower CO abundance in the outer envelope.

Following \cite{Justtanont2005}, another option for tacklng the problem is to vary the parameter that controls the point from
which dissociation of CO happens, r$_{1/2}$. We calculated new models considering values r$_{1/2}$=0.2, 0.3, 0.4, 0.5, and 0.8 r$_{1/2}^{\rm \,m}$.
This corresponds to a stronger dissociation than a value of unity, which corresponds to the prediction of \cite{Mamon1988}, and, therefore, a smaller CO radius.
For r$_{1/2}$= r$_{1/2}^{\rm \,m}$, the CO abundance has decreased by half at $\sim$\,860 stellar radii.
This is at about the region where the PACS images show that there is more dust emission than predicted by our dust model.
By decreasing r$_{1/2}$ to 0.4 r$_{1/2}^{\rm \,m}$, which corresponds to CO reaching half of the initial abundance around 350 stellar radii, being 40\% of the size predicted by \citep{Mamon1988}, we could fit the shape of
transitions $^{12}$CO $J$=4-3, $^{12}$CO $J$=3-2, and $^{12}$CO $J$=2-1 and the total flux of transitions $^{12}$CO $J$=3-2 and $^{12}$CO $J$=2-1. The strength of transition
$^{12}$CO $J$=4-3 is still over-predicted by 60\% by this new model, and the number of CO molecules in the $J$=1 level becomes so small that
the $^{12}$CO $J$=1-0 transition is predicted to be too weak to observe.

Because the PACS image shows an unexpectedly strong dust emission at the radii where we expect the $J$=1-0 line of $^{12}$CO to be excited (the $J$=1
level population reaches its maximum at 1000 R$_{\star}$ in the absence of photo-dissociation),
we do not attempt to model this emission further  in the context of our model assumptions.

\begin{figure}[h]
   \centering
   \includegraphics[width= 8.5cm]{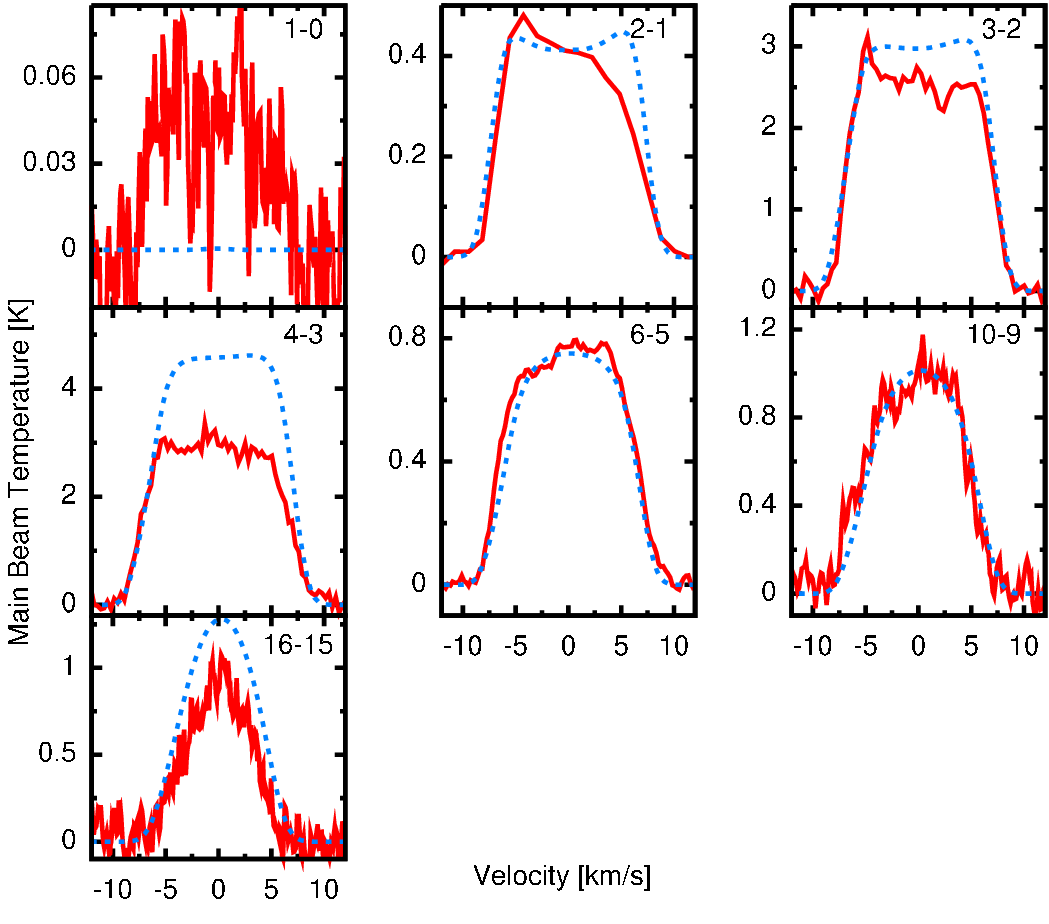}
      \caption{Model with r$_{1/2} = 0.4$, $\beta=5.0,$ and \mdot = 1.3 $\times 10^{-7}$ M$_{\odot}$yr$^{-1}$
      for $^{12}$CO (in dashed blue) compared to observed line shapes (in solid red). Observations were carried out by
      HIFI, APEX, SMT, and SEST (see Table \ref{table:WHya_line_fluxes}).}
         \label{fig:best_fit_shape_smaller_radius}
   \end{figure}

The fit using r$_{1/2} = 0.4$ does not affect the flux of $J$=8-7 and higher excitation lines, because those are formed deep
inside the envelope. The model with smaller CO radius and $\beta$=5.0 is compared to the observed $^{12}$CO line shapes in Fig. \ref{fig:best_fit_shape_smaller_radius}.

\begin{figure}[h!]
\centering
\includegraphics[width=8cm]{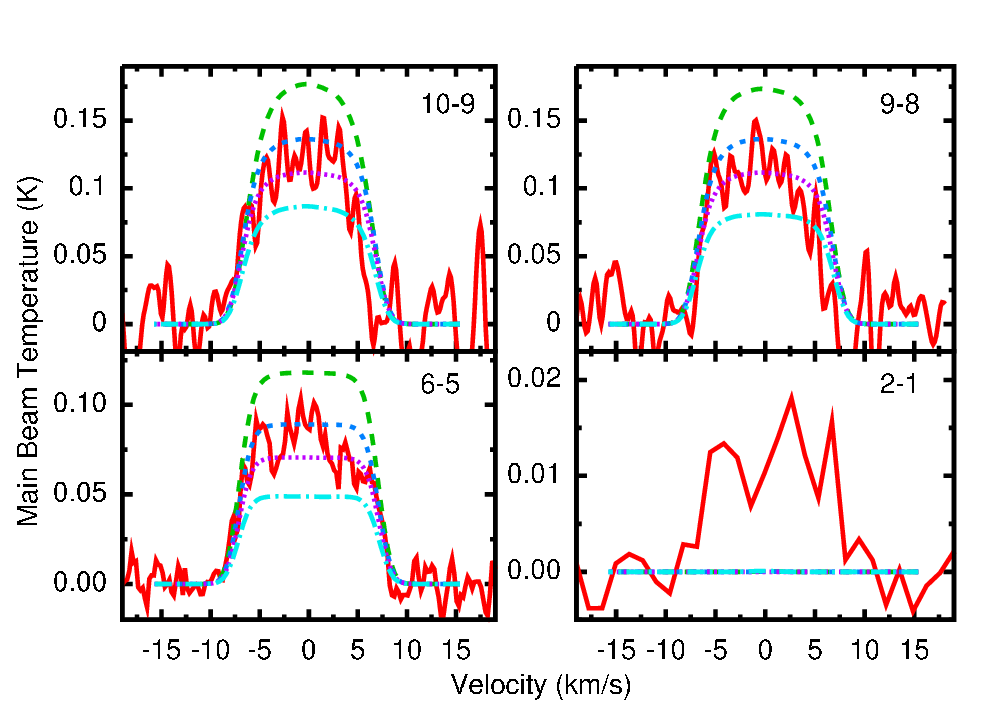}
\caption{Models with r$_{1/2} = 0.4$, $\beta$ = 5.0, and \mdot = 1.3 $\times 10^{-7}$ M$_{\odot}$yr$^{-1}$ compared to the $^{13}$CO observed lines. The solid red line represents the observations.
The long-dashed green, short-dashed blue, dotted pink, and dot-dashed light blue represent models with $^{12}$C/$^{13}$C = 10,15, 20, and 30, respectively. A value of 18 for the $^{12}$C/$^{13}$C
ratio reproduces the data best.}
\label{fig:13CO_smaller_radius}
\end{figure}

In Fig. \ref{fig:13CO_smaller_radius} we show models for the $^{13}$CO transitions, considering the smaller dissociation radius and $\beta$ = 5.0. The
$^{13}$CO $J$=10-9, $^{13}$CO $J$=9-8, and $^{13}$CO $J$=6-5 lines are now more consistently fitted by an isotopic ratio between 15 and 20.
In Sect. \ref{sec:disc} we discuss the uncertainty on the determined isotopic ratio.
The predicted $^{13}$CO $J$=2-1 emission is so weak that it should not be measurable, similar to $^{12}$CO $J$=1-0. Since the population of level $J$=2 of $^{13}$CO peaks even
further out than $J$=1 of $^{12}$CO, we also refrain from explaining the emission in this line in the context of our current model.

\section{Discussion}
\label{sec:disc}

\subsection{The outer CO envelope}

The modelling of W\,Hya suggests a wind structure that is atypical of oxygen-rich AGB sources, both in terms of the behaviour of 
the CO gas and of the dust, particularly so near the CO photodissociation zone.  Using maps of 
$^{12}$CO $J$=2-1 and $^{12}$CO $J$=1-0 for other sources, \citet{Castro-Carrizo2010} show that the location of this zone matches, or is larger than, the photodissociation radius
predicted by \cite{Mamon1988}.  The strengths and shapes of low-excitation lines ($J_{\rm up} \le 4$)
in W\,Hya,  however, clearly indicate that the size of the CO envelope is significantly smaller than predicted by \cite{Mamon1988}. 
That the situation is complex may be discerned from the $^{12}$CO $J$=1-0 and $^{13}$CO $J$=2-1 pure rotational lines. Weak emission is observed 
for these lines, but no emission is predicted when we adopt r$_{1/2}$ = 0.4 r$_{1/2}^{\rm \,m}$. These lines are, however, mainly excited
in the outer parts of the CO envelope (r $\gtrsim$ 600 R$_{\star}$, or 15$\arcsec$ for the adopted stellar parameters and distance),
a region where the PACS images show that the envelope of W\,Hya is not  predicted well by our constant mass-loss rate model.

\subsection{The $^{12}$CO/$^{13}$CO ratio}

For the first time, we have multiple isotopic transitions in $^{12}$CO and $^{13}$CO available to constrain the $^{12}$CO/$^{13}$CO 
isotopologic ratio.  Three of the four lines for which we have data for the $^{13}$CO isotopologue, i.e. $J$=10-9, 9-8, and 6-5, point to an intrinsic 
$^{12}$CO/$^{13}$CO ratio of 18. For comparison, the line ratios between the two isotopologues observed by HIFI are around 8 and 8.7 for transitions
$J$=10-9 and $J$=6-5, respectively.
The only previous attempt to constrain the
intrinsic ratio is  by \cite{Milam2009}, who found a value of 35 using the $J$=2-1 of $^{13}$CO transition. We could not use
this particular transition, since it is poorly reproduced by our model.
We note that \citeauthor{Milam2009} considered a mass loss that is an order of magnitude greater than was found by us and \cite{Justtanont2005}, 
and a three times lower $^{12}$CO abundance. They also assumed W\,Hya to be at a distance of 115 pc, which
implies a twice higher luminosity than we have adopted. Given the intricacies in forming the $^{12}$CO and $^{13}$CO
lines, especially in the case of the outer envelope of W\,Hya, it is hard to assess the meaning of the factor-of-two difference in the intrinsic $^{12}$C/$^{13}$C.
We therefore discuss the uncertainties in this ratio in more detail.  This discussion is also important in view of a comparison with
theoretical predictions, this ratio being a very useful tool for constraining AGB evolution models.

\subsubsection{The uncertainty of the determined $^{12}$CO/$^{13}$CO ratio}

In order to quantify the robustness of the derived $^{12}$CO/$^{13}$CO ratio, we carried out a test in which we vary parameters
that so far were held fixed in our model grid and studied the effect.

First, we investigate the impact of varying 
the assumed $^{12}$CO abundance relative to H (standard at $2.0 \times 10^{-4}$) and the stellar luminosity considered (5400\,$L_{\odot}$).
We calculated models in which we changed each of these parameters, but only one at a time, by a factor of two -- both toward higher and lower values.
When these parameters are varied, the $^{12}$CO line fluxes and shapes also change. We scaled the mass-loss rate in order to fit the line 
flux observed by SPIRE, since the $^{12}$CO lines observed by this instrument are excited in the same region as $^{13}$CO transitions used
to determine the isotopic ratio. We did not attempt a fine-tuning to match the strengths and shapes of the profiles. 
The factor-of-two changes in each of the two parameters affect the line fluxes of $^{13}$CO by typically 20 to 25 percent and at most 
30 percent.

Second, we studied the effect of changes in the input stellar spectrum. As pointed out in Sect. \ref{sec:radiation_field}, the stellar flux in the near infrared
may have an important effect on the CO line strengths as  photons of these wavelengths can efficiently pump molecules to higher rotational levels.  
In our models, the spectral emission of W\,Hya is approximated by a black body of 2500 K. However, the stellar spectrum is much more complex, and its intensity
changes considerably even for small differences in wavelength. Specifically, molecules in the stellar atmosphere can absorb photons, thereby reducing
the amount available for exciting $^{12}$CO and $^{13}$CO further out in the envelope. Figure \ref{fig:fit_ISO} indicates that the molecular absorption at
4.7 $\mu$m is less than a factor of two. We assume a factor-of-two decrease in the stellar flux in the 4.7 $\mu$m region to assess the impact of the near-IR flux on the modelled lines.
The change in the integrated line flux of $^{12}$CO lines is only about five percent. The $^{13}$CO lines respond more strongly, varying by typically 25\%.

The combined impact of these three sources of uncertainties on the derived isotopic ratio would be $\sim$\,40\% in this simple approach.
Also accounting for uncertainties due to the flux calibration and noise (which are about 25\% uncertain) and the actual model fitting
(25\% uncertain), we conclude that the intrinsic isotopic ratio determined here is uncertain by 50\% to 60\%.  This implies that
W\,Hya has a $^{12}$CO/$^{13}$CO ratio of $18 \pm 10$. 
Better signal-to-noise observations of, particularly, $^{13}$CO, in combination with
spatial maps of both $^{12}$CO and $^{13}$CO lines, are required to constrain this value better.

\subsubsection{Connecting the $^{12}$C/$^{13}$C ratio to stellar evolution}

From evolutionary model calculations, the surface $^{12}$C/$^{13}$C ratio is expected to change in dredge-up events. It is found to decrease by a factor of a few after the first dredge-up,
taking place during the ascent on the red giant branch (RGB) and reaching a value of typically 20 for stars more massive than 2 M$_\odot$. For stars with masses lower than 2 M$_\odot$, it is found to
increase with decreasing mass, reaching about 30 at 0.8 M$_\odot$.
In stars that experience the second dredge-up (M$_\star \gtrsim$ 4.5 M$_\odot$), this ratio is found to decrease further by a few tens of percent during the ascent on the AGB
\citep{Busso1999,Karakas2011}. Stars with masses higher than 1.5 M$_\odot$ experience the third dredge-up, a continuous process that occurs during the thermally pulsing AGB phase. Evolutionary
models show that this ratio is steadily increasing thanks to the surface enrichment of $^{12}$C or is not changing significantly if the star is massive enough, M$_\star \gtrsim 4.0$ M$_\odot$, for hot bottom burning to operate.

Recent calculations include extra-mixing processes to explain the low values of the $^{12}$C/$^{13}$C
ratio observed in low mass RGB stars \citep{Tsuji2007,Smiljanic2009,Mikolaitis2012}. Extra-mixing processes are thought to occur
also during the AGB phase \citep{Busso2010}, but its causes and consequences are more uncertain.
Models that include extra mixing in the RGB predict lower isotopic ratios ($\sim$\,10) for low mass stars ($\sim$\,1\,M$_\odot$).

When compared to model predictions, a value of 18 for the $^{12}$C/$^{13}$C ratio would be consistent with W\,Hya having an isotopic ratio that reflects the value set by the
first dredge-up \citep{Boothroyd1999,Charbonnel2010} for star with masses higher than about 2 M$_\odot$. 
 If W\,Hya's mass lies between 1.5 M$_\odot$ and 4.0 M$_\odot$, the value found by us further suggests that the star has experienced few or none of these third dredge-up events.

Our intrinsic isotopic ratio, however, is not very constraining, since it agrees within the uncertainties with three very different scenarios for W\,Hya's evolutionary stage: first, having a
low value of this ratio ($\sim$\,8), hence having suffered extra mixing in the first dredge-up, characteristic of low mass stars; second,  having a ratio of indeed 18, which implies
a higher mass; or, third, having a higher ratio ($\sim$\,28) and being on its way to becoming a carbon star.
Decreasing the error bars by a factor of three or four would allow one to draw stronger conclusions on this matter.
Such accuracy may be achieved with the Atacama Large Millimeter/submillimeter Array.

\subsection{The wind acceleration}

The wind acceleration in W\,Hya is quite slow.
The highest excitation line for which we have an observed line shape,  $^{12}$CO $J$=16-15, has an excitation of its upper level that peaks at eight stellar radii, decreasing to one-fifth of this
peak value at 30\,$R_{\star}$.  The triangular shape and the width of this profile indicate that it is formed in the accelerating part of
the flow.  The $^{12}$CO $J$=10-9 line is still explained well by a $\beta$ = 5.0 model, but the region where this line is excited seems to be where the wind starts to be
accelerated faster than our model with $\beta$ = 5.0. Interestingly, this effect is noted mainly in the blue-shifted part of the flow. The population of level $J$=10 of $^{12}$CO reaches a maximum at
25 $R_{\star}$, decreasing to one-fifth of this peak value at 70 $R_{\star}$. This shows that the wind approaches the terminal velocity indeed much later than expected,
somewhere around 50 stellar radii. Furthermore, transition $J$=6-5 of $^{12}$CO is formed in a region where the wind has already reached maximum
expansion velocity, contrary to what a model with $\beta$=5.0 predicts. This indicates that, although the wind has a slow start until 5.5 - 6.0 km/s, the last
injection of momentum happens quite fast. Other authors have also concluded that the wind of W\,Hya is accelerated slower than expected
\citep{Lucas1992,Szymczak1998}, in agreement with our results.

The shapes of $^{13}$CO $J$=10-9 and $J$=9-8 lines are also asymmetric, as is that of $^{12}$CO $J$=10-9 line. The reason for this is not clear but
may be connected to large scale inhomogeneities that damp out, or smooth out, at large 
distances. A direction-dependent acceleration law, for example, or direction-dependent excitation structures of the higher levels of these two transitions
might be the reason we see this asymmetry. The higher optical depth in the $^{12}$CO lines might be able to make this feature less
pronounced in the $J$=10-9 of this isotopologue. We note that these two $^{13}$CO lines do not have a high signal-to-noise ratio, therefore, no firm conclusion can be drawn
based on this apparent asymmetry.

\section{Summary}
\label{sec:summary}

We have constrained the wind structure of W\,Hya using an unprecedented number of $^{12}$CO and $^{13}$CO emission lines.
We were especially  interested in understanding the excitation of $^{12}$CO and $^{13}$CO for this source.
The envelope structure derived in this study will enable analysis of other molecular abundances in the outflow, such as ortho- and para-water and its isotopologues,
SiO and its isotopologues, SO, SO$_2,$ and even carbon-based molecules such as HCN.  Specifically, we may thus obtain excitation conditions of these molecules and the
heating and cooling rates -- mainly thanks to water transitions -- associated to it. These species too will add to our understanding of the physical and chemical processes in the wind. 

The main conclusions obtained from modelling $^{12}$CO and $^{13}$CO are

\begin{itemize}

\item{The model that best fits the data has a mass-loss rate of $1.3 \times 10^{-7}\ M_{\odot}$yr$^{-1}$, an expansion velocity of $7.5$ km/s,
a temperature power-law exponent of $0.65$, a CO dissociation radius 2.5 times smaller than what is predicted by theory, and an exponent of the $\beta$-type velocity law of 5.0.
We note that the wind has a slow start that is better reproduced by a high value of this exponent, but that the envelope reaches its final expansion velocity sooner than such a model would predict.}

\item{The smaller outer CO radius is supported mainly by the line strengths of the low-$J$ lines. Introducing a broken temperature law does not fix this
problem, and a varying mass-loss rate, lower in the outer envelope, seems to contradict what is seen in the PACS dust maps.}

\item{By comparing our constant mass-loss rate dust model with recently published PACS images of W\,Hya, we note that our dust model does not reproduce
the observations beyond 20$\arcsec$, corresponding to 800 R$_\star$ for the adopted parameters and distance. This extra emission may originate in material expelled in a phase of higher mass loss
or be the result of a build up of material from interaction with previously ejected gas or interstellar medium gas.}

\item{We derive a $^{12}$CO to $^{13}$CO isotopic ratio of 18 $\pm$10. The accuracy is not sufficient to draw firm conclusions on the evolutionary stage or main-sequence mass
of W\,Hya, but a ratio of 20 would be expected for an AGB star with mass higher than 2 M$_\odot$ that did not experience $^{12}$C enrichment due to the third dredge-up phase.
Spatially resolved observations may help constrain the $^{12}$CO abundance and the $^{13}$CO excitation
region and allow for a more precise estimate of this ratio.}

\end{itemize}
 
 \begin{acknowledgements}
HIFI has been designed and built by a consortium of
institutes and university departments from across Europe, Canada, and the
United States under the leadership of SRON Netherlands Institute for Space
Research, Groningen, The Netherlands and with major contributions from
Germany, France, and the US. Consortium members are Canada: CSA,
U. Waterloo; France: CESR, LAB, LERMA, IRAM; Germany: KOSMA,
MPIfR, MPS; Ireland, NUI Maynooth; Italy: ASI, IFSI-INAF, Osservatorio
Astrofisico di Arcetri-INAF; Netherlands: SRON, TUD; Poland: CAMK, CBK;
Spain: Observatorio Astron\'omico Nacional (IGN), Centro de Astrobiolog\'{i}a
(CSIC-INTA). Sweden: Chalmers University of Technology Ð MC2, RSS \&
GARD; Onsala Space Observatory; Swedish National Space Board, Stockholm
University Ð SStockholm Observatory; Switzerland: ETH Zurich, FHNW; USA:
Caltech, JPL, NHSC.
PACS has been developed by a consortium of institutes
led by MPE (Germany) and including UVIE (Austria); KUL, CSL,
IMEC (Belgium); CEA, OAMP (France); MPIA (Germany); IFSI, OAP/AOT,
OAA/CAISMI, LENS, SISSA (Italy); IAC (Spain). This development has been
supported by the funding agencies BMVIT (Austria), ESA-PRODEX (Belgium),
CEA/CNES (France), DLR (Germany), ASI (Italy), and CICYT/MCYT (Spain).
SPIRE has been developed by a consortium of institutes led by Cardiff Univ. (UK) and
including Univ. Lethbridge (Canada); NAOC (China); CEA, LAM (France);
IFSI, Univ. Padua (Italy); IAC (Spain); Stockholm Observatory (Sweden);
Imperial College London, RAL, UCL-MSSL, UKATC, Univ. Sussex (UK);
Caltech, JPL, NHSC, Univ. Colorado (USA). This development has been supported
by national funding agencies: CSA (Canada); NAOC (China); CEA,
CNES, CNRS (France); ASI (Italy); MCINN (Spain); SNSB (Sweden); STFC
(UK); and NASA (USA).
T.Kh. gratefully acknowledges the support from NWO grant 614.000.903.
R.Sz. and M.Sch. acknowledge support from NCN grant N 203 581040.
This work has been partially supported by the Spanish MICINN, programme
CONSOLIDER INGENIO 2010, grant "ASTROMOL" (CSD2009-00038).
JB, PR, BvB acknowledge support from the Belgian Science Policy office
through the ESA PRODEX programme. FK is supported by the FWF project
P23586 and the  ffg ASAP project HIL.
\end{acknowledgements}

\bibliographystyle{aa}

\end{document}